\documentclass[aps,pre,twocolumn,groupedaddress, amsmath, amssymb]{revtex4-2}

\usepackage{graphicx} 
\usepackage{amsmath} 
\usepackage{tikz} 
\usetikzlibrary{shapes}
\usetikzlibrary{patterns}
\usetikzlibrary{angles, quotes}

\usepackage[colorinlistoftodos, draft]{todonotes}

\usepackage{amssymb}
\usepackage{bbold}

\begin{document}


\let\vvec\vec
\renewcommand{\vec}[1]{\mathrm{\mathbf{#1}}}
\newcommand{\unit}{\vec{e}}

\newcommand{\rhotwo}{\rho^{(2)}}
\newcommand{\rv}{{\bf r}}
\newcommand{\xv}{{\bf x}}
\newcommand*\colvec[3][]{\begin{pmatrix}\ifx\relax#1\relax\else#1\\\fi#2\\#3\end{pmatrix}}

\title{Fundamental measure theory of inhomogeneous two-body correlation functions}


\author{S.~M. Tschopp}
\author{J.~M. Brader}
\affiliation{Department of Physics, University of Fribourg, CH-1700 Fribourg, Switzerland}



\begin{abstract}
For the three-dimensional hard-sphere model we investigate the inhomogeneous two-body 
correlations predicted by Rosenfeld's fundamental measure theory.  
For the special cases in which the density has either planar or spherical symmetry 
we provide analytic formulae for the Hankel and Legendre transforms, respectively, of the 
inhomogeneous two-body direct correlation function as explicit functionals of the density.
When combined with the Ornstein-Zernike relation our analytical results allow for 
rapid calculation of inhomogeneous hard-sphere density correlations in real-space. 
These provide not only information about the packing structures of the hard-sphere 
system, but also form 
an essential building-block for constructing perturbation theories of more realistic 
models.   
\end{abstract}


\maketitle



\section{Introduction}


Two-body correlation functions give important information about the microstructural particle 
arrangement in a classical fluid. 
In the presence of an external field the density becomes nonuniform and the corresponding 
inhomogeneous two-body correlations can deviate significantly from those in bulk, 
e.g.~for fluids at interfaces or under spatial confinement.
For systems interacting via a pair potential these deviations provide direct access to 
interfacial thermodynamic quantities, such as the surface tension between coexisting phases 
\cite{evans79,fundamentals,Widom}. 

The most familiar theoretical approach to calculating two-body correlations is the method of integral 
equations, based on closures of the Ornstein-Zernike (OZ) equation \cite{Hansen06}. 
Approximations such as the Percus-Yevick (PY) or the hypernetted-chain 
have been widely used to study thermodynamics and two-body correlations in bulk \cite{caccamo}, where  
translational invariance enables fast Fourier transform methods to be employed to great 
advantage in numerical calculations. 
For inhomogeneous systems this luxury is absent; the two-body correlation functions 
generally depend upon two vector arguments. 
However, for systems in which the density has a simple geometry (usually planar or spherical) 
the OZ equation can be reduced to a more manageable form. 
In such cases, generalizations of the bulk closure approximations have been used 
to calculate the inhomogeneous two-body correlations (see Refs.\cite{kjellander1,attard1,kjellander2,
henderson_sokolowski,kovalenko,brader2008,botan,nygard1,nygard2}
for examples and chapter four in Ref.\cite{fundamentals} for an overview).

An alternative approach is to use classical density functional theory (DFT). 
Within the DFT framework, correlation functions are generated by successive functional differentiation 
of the excess Helmholtz free energy functional with respect to the density. 
The functional contains complete statistical information about the system and can thus be used to calculate correlation 
functions of any order. 
Calculation of the inhomogeneous two-body correlations proceeds in the following way: 
(i) minimize the grand potential functional to obtain the equilibrium one-body density,  
(ii) evaluate the two-body direct correlation function (generated by taking two functional derivatives of the excess 
Helmholtz free energy) at the 
equilibrium density and then solve the OZ equation for the two-body total correlation function. 
No closure is required, as the direct correlation function is uniquely 
specified by the generating functional. 
This two-step scheme, sometimes referred to as the `Ornstein-Zernike route' 
is often used to obtain {\it bulk} two-body correlations (in which case the equilibrium density is a trivial 
constant), but is more rarely exploited to address inhomogeneous systems 
(for some examples see, 
e.g.~Refs.\cite{dietrich1,dietrich2,dietrich3}).

The most well-studied model in liquid-state theory is the hard-sphere system. 
In a classic 1989 paper Rosenfeld introduced a geometrically-based fundamental measure theory (FMT)
density functional for hard-spheres \cite{rosenfeld89}. 
The predictions of Rosenfeld FMT for the one-body density profile were found to be in excellent agreement with 
computer simulation data for a wide variety of external fields \cite{roth2010}. 
Although the original FMT encountered difficulties for strongly confined fluids and ordered states, subsequent 
versions of hard-sphere FMT corrected these shortcomings. 
The FMT, in common with other DFT approximations, is usually employed to obtain the equilibrium one-body density 
profile in a given external potential. 
Higher-body correlation functions are typically only evaluated in bulk \cite{rosenfeld89} and 
{\it inhomogeneous} pair and higher-order correlation functions from FMT remain largely 
unexplored. 
This is perhaps surprising, given that the analytic formulae for the direct correlation functions present 
an obvious (and computationally advantageous) alternative to the inhomogeneous integral equation closures 
mentioned above. 
A deeper investigation of higher-order FMT correlation functions 
would not only provide insight into the structure of hard-sphere FMT, possibly suggesting improvements, 
but is also needed for the construction of perturbation theories aiming to describe more realistic inhomogeneous fluids.   

In this paper we will address these issues and analyze in detail the inhomogeneous two-body correlations 
generated by FMT. 
We focus on situations for which the one-body density exhibits either planar or spherical 
symmetry and derive analytic formulae for the Hankel (planar geometry) and
Legendre (spherical geometry) transforms of the inhomogeneous two-body FMT direct correlation function. 
These explicit functionals of the (one-dimensional) density profiles then provide rapid access to the 
direct and total pair correlation functions in real-space. 
Our results for hard-spheres will be tested against the inhomogeneous PY integral equation theory and 
existing Monte-Carlo data. 
Once the quality of the FMT correlations has thus been established we will show how these can 
be exploited as input to a recently developed perturbative density functional theory for treating 
systems with attractive interactions \cite{tschopp1}.  

The paper will be structured as follows: In Section \ref{theory} we briefly outline relevant aspects of classical DFT. 
In Section \ref{theoryHS} we introduce the 
FMT and give explicit formulae for the Hankel and Legendre transforms of the two-body 
direct correlation function in planar and spherical geometry, respectively. 
In Section \ref{results_hs} we present numerical results for the inhomogeneous total correlation function of 
hard-spheres confined between planar walls and in the presence of a fixed test particle. 
In Section \ref{theorybh} we show how our results for hard-spheres can be used as input to a 
perturbation theory of attractive interactions. 
In Section \ref{resultsbh} we give results obtained using this perturbation theory for 
the well-known hard-core Yukawa model. 
Finally, in Section \ref{discussion}, we discuss our findings and give an outlook for future 
investigations.

\section{Density functional theory}\label{theory}
DFT is an exact formalism for the study of 
classical many-body systems in external fields 
\cite{evans79,fundamentals,Widom}. 
The central object of interest is the grand potential functional
\begin{align}\label{grand}
\Omega[\,\rho\,] = F^{\rm id}[\,\rho\,] + F^{\rm exc}[\,\rho\,] 
- \int \!d\rv \big( \mu - V_{\rm ext}(\rv) \big)\rho(\rv), 
\end{align}
where $\mu$ is the chemical potential, $V_{\rm ext}(\rv)$ is the external potential and 
$\rho(\rv)$ is the one-body ensemble averaged density. 
The square brackets indicate a functional dependence.
The Helmholtz free energy of the ideal gas is exactly given by	
\begin{eqnarray}\label{idealfree}
F^{\rm id}[\,\rho\,]=k_BT\int\!d\rv\, \rho(\rv)\left(\, \ln(\rho(\rv))-1\, \right), 
\end{eqnarray}
where $k_B$ is the Boltzmann constant, $T$ is the temperature and 
we have set the thermal wavelength equal to unity. 
The excess Helmholtz free energy, $F^{\rm exc}$, includes all information 
regarding the interparticle interactions and usually has to be approximated.
The grand potential satisfies the variational condition
\begin{align}
\label{EQomegaMinimial}
\frac{\delta  \Omega[\rho\,]}{\delta \rho(\rv)}=0. 
\end{align}
This yields the following Euler-Lagrange equation for the equilibrium 
one-body density
\begin{align}\label{euler}
\rho(\rv)=e^{ -\beta\left(V_{\rm ext}(\rv) - \mu - k_BTc^{(1)}(\rv)\right)},
\end{align}
where the one-body direct correlation function is generated from the excess 
Helmholtz free energy by a functional derivative
\begin{align}\label{1derivative}
c^{(1)}(\rv)=-\frac{\delta 
\beta F^{\rm exc}}{\delta\rho(\rv)}.
\end{align}
Substitution of the solution of \eqref{euler} into \eqref{grand} yields the equilibrium 
grand potential, thus providing access to all thermodynamic properties of 
the system. 

Information about the two-body correlations in the inhomogeneous fluid can be obtained 
from a second functional derivative of the free energy  
\begin{align}\label{2derivative}
c^{(2)}(\rv_1,\rv_2)=-\frac{\delta^2 
\beta F^{\rm exc}}{\delta\rho(\rv_1)\delta\rho(\rv_2)},
\end{align}
where $c^{(2)}$ is the two-body direct correlation function. 
The connection between $c^{(2)}$ and the total correlation function, $h$,  
can be established by considering the 
functional derivative of equation \eqref{EQomegaMinimial} with respect to the 
external field 
\begin{align}\label{doublederiv}
\frac{\delta^2  \Omega}{\delta V_{\rm ext}(\rv_1)\delta \rho(\rv_2)}=0.   
\end{align}
While the vanishing of this mixed derivative is a trivial consequence of 
\eqref{EQomegaMinimial}, it is nevertheless a useful result. 
Explicit calculation of the left-hand side
of \eqref{doublederiv} yields the OZ equation
\begin{align}\label{oz}
h^{}(\rv_1,\rv_2) = c^{(2)}(\rv_1,\rv_2) \!+\! 
\int\! d\rv_3\, h^{}(\rv_1,\rv_3)\rho(\rv_3) c^{(2)}(\rv_3,\rv_2),  
\end{align} 
%
Note that the external potential does not appear explicitly in the OZ equation, this information 
is implicitly contained within the one-body density.

\section{Hard-spheres}\label{theoryHS}

\subsection*{Fundamental measure theory}

We now focus our attention on a system of three-dimensional hard-spheres of 
radius $R$. 
Within FMT the excess Helmholtz free energy is approximated by an integral over a 
function of weighted densities \cite{rosenfeld89}
\begin{align}\label{ros_fe}
\beta F^{\rm exc}_{\rm hs}
[\,\rho\,] = \int d\rv_1 \; \Phi \left( \left\lbrace n_{\alpha}(\rv_1) \right\rbrace  \right). 
\end{align}
The original Rosenfeld formulation of FMT employs the following reduced excess free energy density 
\begin{equation}\label{Phi_ros_fe}
\Phi = - n_0 \ln(1-n_3) + \frac{n_1 n_2 - {\bf n}_1 \cdot {\bf n}_2}{1-n_3} + \frac{n_2^3 
- 3 n_2 {\bf n}_2 \cdot {\bf n}_2}{24 \pi (1-n_3)^2}.
\end{equation}
The weighted densities are generated by convolution	
\begin{equation}\label{n_alpha_ros_fe}
n_{\alpha}(\rv_1) = \int d\rv_2 \; \rho(\rv_2)\, \omega_{\alpha}(\rv_1-\rv_2), 
\end{equation}
where the weight functions, $\omega_{\alpha}$, are characteristic of the geometry of the spheres. 
Of the six weight functions, four are scalars 
\begin{align}
\omega_3(\rv)&=\Theta(R-r), \hspace*{0.5cm}
\omega_2(\rv)=\delta(R-r), \notag\\
\omega_1(\rv)&=\frac{\delta(R-r)}{4\pi R}, \hspace*{0.51cm}
\omega_0(\rv)=\frac{\delta(R-r)}{4\pi R^2}, \notag
\end{align}
and two are vectors
\begin{align}
\omega_{\bold 2}(\rv)&=\unit_{\rv}\,\delta(R-r),\hspace*{0.5cm}
\omega_{\bold 1}(\rv)=\unit_{\rv}\frac{\delta(R-r)}{4\pi R},
\notag
\end{align}
where $\unit_{\rv}=\rv/r$ is a unit vector. 
The presence of summations over both vector and scalar weights in many FMT expressions 
presents some notational difficulty and the analytical calculations below demand clarity regarding 
the scalar or vector character of various functions. 
We have thus chosen to employ the symbol $\omega$ for all weight functions, both scalar and vector, 
where the latter will be distinguished by employing a bold font index. 
This choice also enables us to use the convenient notation $\omega_{|{\bold 2}|}\!=\!\omega_{2}$ 
and $\omega_{|{\bold 1}|}\!=\!\omega_{1}$.  

Applying the definition \eqref{1derivative} to the free energy \eqref{ros_fe} generates 
the following approximate form for the one-body direct correlation function 
\begin{align}\label{c_onebody}
c^{(1)}_{\rm hs}(\rv_1)=-\sum_{\alpha}\int d\rv_2\, 
\Phi^{'}_{\alpha}(\rv_2)\,\omega_{\alpha}(\rv_{21}),
\end{align}
where $\Phi^{'}_{\alpha}\!=\!\partial \Phi/\partial n_{\alpha}$, $\rv_{21}=\rv_2-\rv_1$ 
and the summation runs over all scalar and vector indices. 
Consistent with our established notation the function $\Phi'_{\alpha}$ 
is a vector quantity when $\alpha$ takes the value ${\bold 1}$ or ${\bold 2}$, in which case 
a scalar product is implied in equation \eqref{c_onebody}. 
While equation \eqref{c_onebody} makes an appearance in practically all FMT studies, 
the FMT approximation to the two-body direct correlation function is less frequently encountered 
and its general structure thus deserves some careful attention. 
Taking two functional derivatives of the free energy \eqref{ros_fe} generates the following 
expression
\begin{align}
c^{(2)}_{\rm hs}(\rv_1, \rv_2)
&=-\!\sum_{\alpha\beta}\int \!d\rv_3\, 
\omega_{\alpha}(\rv_{31})\,
\Phi^{''}_{\alpha \beta}(\rv_3)\,
\omega_{\beta}(\rv_{32}),
\notag
\\
&\equiv -\sum_{\alpha\beta} c_{\alpha\beta}(\rv_1, \rv_2)
\label{c_ros_fe}
\end{align}
where $\Phi^{''}_{\alpha \beta}\!=\!\partial^2 \Phi/\partial n_{\alpha} \partial n_{\beta}$. 
The terms, $c_{\alpha\beta}(\rv_1, \rv_2)$, contributing to the sum in \eqref{c_ros_fe}, can be separated into three 
distinct classes according to the values of the $\alpha$ and $\beta$ indices: 
1. both scalars, 2. one scalar and one vector, and 3. both vectors. 
Each of these three classes involves a function $\Phi''_{\alpha\beta}$ of different tensorial rank 
(for convenience all first and second 
derivatives of $\Phi$ are given explicitly in Appendix \ref{appendix_Phi''}). 

For terms belonging to class 1, the products in equation \eqref{c_ros_fe} are self-explanatory, as the second 
derivative $\Phi''_{\alpha\beta}$ and the weight functions are all scalars. 
For class 2 terms, we have one scalar weight, one vector weight and a vectorial second derivative function. 
A scalar product between the vector weight and $\Phi''_{\alpha\beta}$ is thus implied. 
For example, if $\alpha\!=\!2$ and $\beta\!=\!{\bold 2}$ then the corresponding 
term in the sum \eqref{c_ros_fe} is given by
\begin{equation}
c_{2{\bold 2}}(\rv_1, \rv_2)
=\!
\int d\rv_3\, 
\omega_{2}(\rv_{31})\,
\Phi^{''}_{2 {\bold 2}}(\rv_3)\cdot
\omega_{\bold 2}(\rv_{32}).
\notag
\end{equation}
%
For terms in class 3, we have two vector weights and $\Phi''_{\alpha\beta}$ is a second rank 
tensor. 
For example, the term with $\alpha\!=\!{\bold 2}$ and $\beta\!=\!{\bold 2}$ is given by a quadratic form
\begin{equation}
c_{{\bold 2}{\bold 2}}(\rv_1, \rv_2)
=\!
\int d\rv_3\, 
\omega_{\bold 2}(\rv_{31})\cdot
\Phi^{''}_{{\bold 2}{\bold 2}}(\rv_3)\cdot
\omega_{\bold 2}(\rv_{32}). 
\notag
\end{equation}
The Helmholtz free energy density of the Rosenfeld FMT \eqref{Phi_ros_fe} 
is quadratic in the vector weighted densities. 
This has the simplifying consequence that $\Phi''_{\alpha\beta}$ for class 3 terms is proportional 
to the unit tensor. 
We note that this would generally not be the case for FMT approximations involving an 
extended set of weight functions (e.g.~the Tarazona FMT \cite{tarazona_tensors}).

\subsection*{Inhomogeneous Percus-Yevick closure}

An alternative approach to obtaining the inhomogeneous pair correlations is to supplement the OZ equation 
\eqref{oz} by a 
second (usually local) closure relation 
between the pair direct correlation function and the total correlation function, 
and then to solve self-consistently the two coupled equations. 
A closure which is known to work well for hard-spheres is the PY approximation 
\cite{Hansen06,attard_book,attard1}
\begin{align}\label{py}
h^{}_{\rm hs}(\rv_1,\rv_2) &= -1 \;\;\;\text{for}\;  |\rv_1-\rv_2|<2R,\notag\\
c_{\rm hs}^{(2)}(\rv_1,\rv_2) &= 0 \;\;\;\;\;\;\text{for}\; |\rv_1-\rv_2|>2R.
\end{align} 
The first of these relations is the exact `core condition' reflecting the impossibility of hard-sphere overlaps. 
The PY theory can be solved exactly in bulk and yields an expression for the pair direct correlation 
function identical to that generated by the Rosenfeld FMT. 
However, as this agreement occurs only in bulk, care should be taken not to label 
the Rosenfeld FMT as the `PY functional'. 
For inhomogeneous situations the predictions of equation \eqref{c_ros_fe} for any given density profile 
will differ from the solution of the coupled equations \eqref{oz} and \eqref{py}. In particular, 
the FMT expression \eqref{c_ros_fe} will not satisfy exactly the core condition, although it may 
provide a good approximation.  
%
The PY theory has been shown to perform well in a variety of inhomogeneous situations \cite{attard1,brader2008,
kovalenko} 
and we will thus use it as a benchmark for assessing the quality of the pair correlations 
generated by FMT.


\subsection*{FMT in planar geometry}

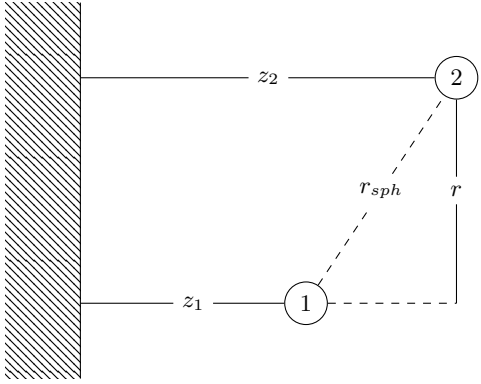
\begin{figure}
\begin{minipage}[t]{0.4\textwidth}
\hspace*{-0.5cm}
\begin{tikzpicture}
\coordinate (cross) at (2,0);
\coordinate (wall upper pt) at (-3,4);
\coordinate (wall lower pt) at (-3,-1);
\coordinate (wall size) at (-1,5);
\node[draw, circle] (particle 1) at (0,0) {1};
\node[draw, circle] (particle 2) at (2,3) {2};
\node (r spherical) at (1,1.5) {$r_{sph}$};
\node (r cylindrical) at (2,1.5) {$r$};
\node (z1 label) at (-1.5,0) {$z_1$};
\node (z2 label) at (-0.5,3) {$z_2$};
\draw[-, >=latex] (-3,0) -- (z1 label);
\draw[-, >=latex] (z1 label) -- (particle 1);
\draw[-, >=latex] (-3,3) -- (z2 label);
\draw[-, >=latex] (z2 label) -- (particle 2);
\draw[dashed, >=latex] (particle 1) -- (r spherical);
\draw[dashed, >=latex] (r spherical) -- (particle 2);
\draw[-, >=latex] (particle 2) -- (r cylindrical);
\draw[-, >=latex] (r cylindrical) -- (cross);
\draw[dashed, >=latex] (particle 1) -- (cross);
\draw[-, >=latex] (wall lower pt) -- (wall upper pt);
\fill[pattern=north west lines] (wall lower pt) rectangle ++(wall size);
\end{tikzpicture}
\end{minipage}
\hspace*{2cm}
\caption{Sketch of the planar geometry.  	
}
\label{planar_sketch}
\end{figure}

When the external field has planar symmetry the density only varies as a function 
of a single cartesian coordinate,  which we take to be the $z$-axis. The inhomogeneous 
pair correlations thus exhibit cylindrical symmetry and require 
as input two coordinates, $z_1$ and $z_2$, and a cylindrical radial distance, $r$, 
separating them (see Fig.\ref{planar_sketch}). 
In this case the OZ equation \eqref{oz} can be simplified using a Hankel transform 
(two dimensional Fourier transform) in the plane orthogonal to $z$. 
The Hankel transform  of the pair direct correlation function is given by 
\begin{align}\label{hankel}
\overline{c}^{\,(2)}_{\rm hs}(z_1,z_2,k) 
= 
2\pi\int_0^{\infty} \!dr\, rJ_0(kr)c^{(2)}_{\rm hs}(z_1,z_2,r),
\end{align}
where $k$ is the absolute value of the two-dimensional wavevector $\vec{k}$ and 
$J_0$ is a Bessel function. 
The back-transform is given by 
\begin{align}\label{hankel_back}
c^{(2)}_{\rm hs}(z_1,z_2,r) 
= 
\frac{1}{2\pi}\int_0^{\infty} \!dk\, kJ_0(kr)\,\overline{c}^{\,(2)}_{\rm hs}(z_1,z_2,k).
\end{align}
Analogous expressions can be written for the total correlation function.
Hankel transform of the OZ equation \eqref{oz} yields (see Appendix \ref{appendix_OZ})
\begin{align}\label{oz_planar}
\overline{h}^{}_{\rm hs}(z_1,z_2,k) &= \overline{c}^{\,(2)}_{\rm hs}(z_1,z_2,k) 
\\
&+ \int_{-\infty}^{\infty} \!dz_3 \;  
\overline{h}^{}_{\rm hs}(z_1,z_3,k)\,\rho(z_3)\,\overline{c}^{\,(2)}_{\rm hs}(z_3,z_2,k). 
\notag
\end{align}
If $\rho$ and $\overline{c}^{\,(2)}_{\rm hs}\!$ are known functions, then 
\eqref{oz_planar} becomes a linear integral equation for the remaining unknown $\overline{h}^{}_{\rm hs}$. 

Equation \eqref{c_ros_fe} gives the general FMT approximation to the pair direct correlation function 
as a functional of the three-dimensional density, but is not in a form suitable for numerical 
implementation. 
This is probably the reason why \eqref{c_ros_fe} has 
not been exploited for the development of liquid-state theory.  
In the following we will show that 
the Hankel transform of equation \eqref{c_ros_fe} can be reduced to an expression which 
allows for rapid and precise numerical evaluation of the pair correlations for any given planar density 
profile. 

This `FMT route' to the hard-sphere pair correlations is computationally efficient for a number of 
reasons:  
(i) The iterative solution of the linear integral 
equation \eqref{oz_planar} is both rapid and stable. 
(ii) The equations can be solved entirely in Hankel space with no need to back-transform to real-space 
during the iteration loop.  
(iii) The inhomogeneous pair correlation functions can be determined for a given value of $k$, 
independently of all other wavevectors. 
Calculations can thus be performed in parallel for different $k$-values. 
It is worth to compare this comfortable situation with the demands of solving numerically 
the nonlinear PY integral equation theory (equations \eqref{oz} and \eqref{py}) where we observe:  
(i) The iterative convergence rate is very slow at high densities and small Broyles mixing parameters must be 
employed to maintain stability \cite{Hansen06,roth2010}.  
(ii) The OZ equation \eqref{oz_planar} is treated in Hankel space, whereas the closure \eqref{py} can 
only be implemented in real-space. This prevents parallel computation and demands an expensive back-and-forth 
Hankel transformation at each iteration step.


Hankel transform of the two-body direct correlation function \eqref{c_ros_fe} generates a sum of terms
\begin{equation}\label{cbar_sum}
\overline{c}^{(2)}_{\rm hs}(z_1, z_2, k) = -\sum_{\alpha\beta} \overline{c}_{\alpha \beta}(z_1, z_2, k).
\end{equation}
The main building blocks for each of the terms in \eqref{cbar_sum} are the Hankel transformed scalar 
weight functions 
\begin{align}
\overline{\omega}_{3}(z_1, z_2, k) &= 2 \pi \frac{R_{12}}{k} \Theta_{12} J_1\left(k R_{12}\right), 
\notag\\
\overline{\omega}_{2}(z_1, z_2, k) &= 2 \pi R \Theta_{12} J_0\left(k R_{12}\right), 
\notag\\
\overline{\omega}_{1}(z_1, z_2, k) &= \frac{\overline{\omega}_{2}(z_1, z_2, k)}{4 \pi R}, 
\notag\\
\overline{\omega}_{0}(z_1, z_2, k) &= \frac{\overline{\omega}_{2}(z_1, z_2, k)}{4 \pi R^2}, 
\label{wH}
\end{align}
where $\Theta_{12}\!\equiv\!\Theta\left(R-|z_{12}|\right)$ is the Heaviside step function, 
$R_{12}^2\!\equiv\!R^2-z_{12}^2$ and $z_{12}=z_1-z_2$.


In the discussion below equation \eqref{c_ros_fe} we identified three classes of terms appearing in the 
sum, grouped according to the values of the pair of indices $\alpha$ and $\beta$. 
For terms belonging to {\bf class 1} the steps involved in transforming 
$c_{\alpha \beta}$ are identical to those required to transform the OZ equation 
(see Appendix \ref{appendix_OZ}). 
This yields for $\alpha,\beta \in \{0, 1, 2, 3\}$ the following one-dimensional integral
\begin{align}\label{class1}
&\overline{c}_{\alpha \beta}(z_1, z_2, k) =\!\int_{a}^{\,b} \!\!dz_3 \;
\overline{\omega}_{\alpha}(z_3, z_1, k) \Phi^{''}_{\alpha \beta}(z_3) 
\,\overline{\omega}_{\beta}(z_3, z_2, k),
\end{align}
for $|z_{12}|\le 2R$ and zero otherwise. 
The integration limits are a consequence of the finite range of the weight functions and are given by 
$a=\max(z_1,z_2)-R$ and $b=\min(z_1,z_2)+R$, respectively.

The mixed terms belonging to {\bf class 2} have one scalar and one vector index. 
For example, if we have $\alpha \in \{0, 1, 2, 3\}$ and $\beta \in \{{\bold 1}, {\bold 2}\}$, 
then we must consider the following scalar product
\begin{align*} 
\Phi^{''}_{\alpha \beta}(z_3)\cdot
\omega_{\beta}(\rv_{32})
=
|\Phi^{''}_{\alpha \beta}(z_3)|\,
\omega_{|\beta|}(\rv_{32})
\,\unit_{z_3}\cdot
\unit^{\,\rm shell}_{32},
\end{align*}
where we define an (outwards pointing) unit vector, orthogonal to the surface of the spherical delta-shell centered at $\rv_2$: 
\begin{align}
\unit^{\,\rm shell}_{32}=
\begin{cases} 
      \frac{\rv_3-\rv_2}{R} & |\rv_3-\rv_2| = R, \vspace*{0.2cm}\\
      \hspace{0.38cm}0 & {\rm otherwise}.
\end{cases}
\end{align}
The scalar product is obtained by simple trigonometry, 
$\unit_{z_3}\!\cdot\,\unit^{\,\rm shell}_{32}=z_{32}/R$.  
As this result depends only on $z_3$ and $z_2$ its presence in the integrand of 
\eqref{c_ros_fe} does not 
interfere with the Hankel transformation and the standard procedure outlined in Appendix \ref{appendix_OZ} 
can be applied without modification. 
We thus obtain the following expression
\begin{align}\label{class2A}
\hspace*{-0.25cm}\overline{c}_{\alpha \beta}(z_1, z_2, k) =
\frac{1}{R}\int_{a}^{\,b} \!dz_3&\;
\overline{\omega}_{\alpha}(z_3, z_1, k)
\,|\Phi^{''}_{\alpha \beta}(z_3)|
\notag\\
&\times z_{32}\;\overline{\omega}_{|\beta|}(z_3, z_2, k),
\end{align}
for $|z_{12}|\le 2R$ and zero otherwise. 
%
The same considerations apply when the rank of the indices is exchanged, 
$\alpha \in \{{\bold 1}, {\bold 2}\}$ and $\beta \in \{0, 1, 2, 3\}$. 
This yields
\begin{align}\label{class2B}
\hspace*{-0.25cm}\overline{c}_{\alpha \beta}(z_1, z_2, k) =
\frac{1}{R}
\int_{a}^{\,b} \!dz_3&\;
\overline{\omega}_{|\alpha|}(z_3, z_1, k)\,z_{31}
\notag\\
&\times\;|\Phi^{''}_{\alpha \beta}(z_3)|\;\overline{\omega}_{\beta}(z_3, z_2, k),
\end{align}
for $|z_{12}|\le 2R$ and zero otherwise.  

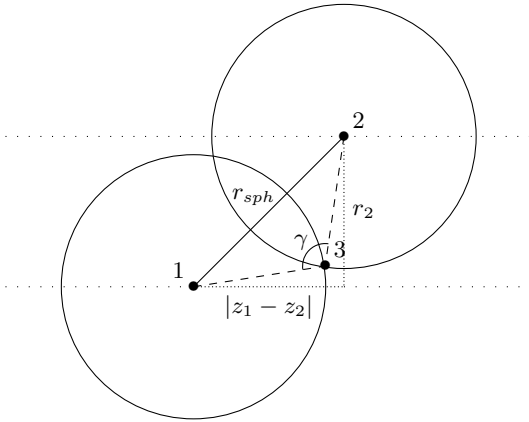
\begin{figure}[!b]
\begin{minipage}[t]{0.4\textwidth}
\hspace*{-0.5cm}
\begin{tikzpicture}
\coordinate (particle 1) at (0,0);
\coordinate (particle 2) at (2,2);
\coordinate (particle 3) at (1.75,0.275);
\coordinate (cross) at (2,0);
\coordinate (z12 label) at (1,0);
\coordinate (r cylindrical) at (2,1);
\coordinate (r spherical) at (0.8,1.2);
\coordinate (upper horizontal left) at (-2.5,0);
\coordinate (upper horizontal right) at (-2.5,2);
\coordinate (lower horizontal left) at (4.5,0);
\coordinate (lower horizontal right) at (4.5,2);
\draw (r spherical) node {$r_{sph}$};
\draw (r cylindrical) node[right] {$r_2$};
\draw (z12 label) node[below] {$|z_1-z_2|$};
\draw (particle 1) node {$\bullet$};
\draw (particle 2) node {$\bullet$};
\draw (particle 3) node {$\bullet$};
\draw (particle 1) node[above left] {$1$};
\draw (particle 2) node[above right] {$2$};
\draw (particle 3) node[above right] {$3$};
\draw[-, >=latex] (particle 1) -- (particle 2);
\draw[densely dotted, >=latex] (particle 2) -- (cross);
\draw[densely dotted, >=latex] (particle 1) -- (cross);
\draw[dashed, >=latex] (particle 1) -- (particle 3);
\draw[dashed, >=latex] (particle 2) -- (particle 3);
\draw (particle 1) circle (50pt);
\draw (particle 2) circle (50pt);
\draw[loosely dotted, >=latex] (upper horizontal left) -- (particle 1);
\draw[loosely dotted, >=latex] (cross) -- (lower horizontal left);
\draw[loosely dotted, >=latex] (upper horizontal right) -- (lower horizontal right);
\pic [draw, -, angle radius=3mm, angle eccentricity=1.5, "$\gamma$"] {angle = particle 2--particle 3--particle 1};
\end{tikzpicture}
\end{minipage}
\hspace*{2cm}
\caption{Geometrical sketch for evaluation of the scalar product given in \eqref{cosgamma}. } 
\label{cosgamma_sketch}
\end{figure}

Terms in {\bf class 3} have a product of two vector weight functions, $\alpha,\beta \in \{{\bf 1}, {\bf 2}\}$, 
and are more difficult to deal with. 
For the original FMT used in this work the second derivative tensor can be expressed as
$\Phi^{''}_{\alpha\beta}\equiv\widetilde{\Phi}^{''}_{\alpha\beta}{\mathbb{1}}$, 
where $\widetilde{\Phi}^{''}_{\alpha\beta}$ is a scalar function and ${\mathbb{1}}$ is the unit tensor.
This enables us to simplify the quadratic form in the integrand of $\overline{c}_{\alpha\beta}$ 
to a scalar product between unit vectors
\begin{align}\label{scalarproduct}
&\omega_{\alpha}(\rv_{31})\cdot
\Phi^{''}_{\alpha\beta}(z_3)\cdot
\omega_{\beta}(\rv_{32})=
\\ 
&
\widetilde{\Phi}^{''}_{\alpha\beta}(z_3)
\,\omega_{|\alpha|}(\rv_{31}) \,\omega_{|\beta|}(\rv_{32})\;
\unit^{\,\rm shell}_{31}
\cdot
\unit^{\,\rm shell}_{32}.
\notag
\end{align}
In Fig.\ref{cosgamma_sketch} we sketch the intersection of a pair of delta-shells centered at $\rv_1$ and $\rv_2$, 
representing a product of weight functions 
$\omega_{\alpha}(\rv_{31}) \,\omega_{\beta}(\rv_{32})$ for $\alpha,\beta\in\{1,2\}$.
The values of the integration variable $\rv_3$ which yield a nonzero contribution to 
\eqref{scalarproduct} lie on the intersection circle of the delta-shells. 
For such cases the points $\rv_1$, $\rv_2$ and $\rv_3$ define an isosceles triangle with fixed angles. 
If we choose $z_1$ as the axis of our cylindrical coordinate system then it is a straightforward 
geometrical exercise to show that for $\rv_3$ anywhere on the intersection circle
\begin{align}\label{cosgamma}
\unit^{\,\rm shell}_{31}
\cdot
\unit^{\,\rm shell}_{32}
\equiv \cos(\gamma)
=1-\frac{z_{12}^2}{2 R^2}-\frac{r_2^2}{2 R^2}, 
\end{align}
where $\gamma$ is defined in Fig.\ref{cosgamma_sketch}. Due to our identification of $z_1$ with the cylindrical 
coordinate axis the variable $r_1$ does not appear in \eqref{cosgamma}. 
We thus seek to evaluate the Hankel transform of  
\begin{align}\label{class3_aim}
c_{\alpha\beta}(z_1,z_2,r_2) = c_{\alpha\beta}^{A}(z_1,z_2,r_2) + c_{\alpha\beta}^{B}(z_1,z_2,r_2)
\end{align} 
where the two contributions are given by 
\begin{align}
c_{\alpha\beta}^{A}(z_1,z_2,r_2)=
\!\!\int\!\! d\rv_3 \widetilde{\Phi}^{''}_{\alpha\beta}(z_3)
\omega_{|\alpha|}(\rv_{31})\, 
\omega_{|\beta|}(\rv_{32})
\!\!\left(
\!1\!-\!\frac{z_{12}^2}{2 R^2}
\!\right)\!,
\notag
\end{align}
\begin{align}
c_{\alpha\beta}^{B}(z_1,z_2,r_2)=
- \!\!\int\!\! d\rv_3 \widetilde{\Phi}^{''}_{\alpha\beta}(z_3)
\omega_{|\alpha|}(\rv_{31})\, 
\omega_{|\beta|}(\rv_{32})
\!\left(
\frac{r_2^2}{2 R^2}
\right)\!.
\notag
\end{align}
The factor $1-z_{12}^2/2R^2$ appearing in the first of these contributions is independent of the radial coordinate.  
The Hankel transformation of $c_{\alpha\beta}^{A}$ thus proceeds in the same way as for the OZ equation 
(see Appendix \ref{appendix_OZ}) and yields
\begin{align}\label{partial_result}
\overline{c}^{\,A}_{\alpha \beta}(z_1, z_2, k) = \left(1-\frac{z_{12}^2}{2 R^2}\right) 
\mathcal A_{\alpha \beta}(z_1, z_2, k),
\end{align}
for $|z_{12}|\le 2R$ and zero otherwise, where the function ${\mathcal A}_{\alpha \beta}$ is given by 
\begin{equation*}
{\mathcal A}_{\alpha \beta}(z_1, z_2, k)\!=\!\! \int_{a}^{\,b}\!\!\!dz_3\, 
\widetilde{\Phi}^{''}_{\alpha \beta}(z_3) \overline{\omega}_{|\alpha|}(z_3, z_1, k) 
\,\overline{\omega}_{|\beta|}(z_3, z_2, k).
\end{equation*}
Hankel transform of $c_{\alpha\beta}^{B}$ is complicated by the presence of the factor $r_2^2$. 
Following again the procedure outlined in Appendix \ref{appendix_OZ}, we find that the first step of the calculation 
can be carried through easily, leading to
\begin{align}\label{tricky}
c^{\,B}_{\alpha \beta}(z_1,z_2,r_2)&=
- \frac{r_2^2}{4\pi R^2}\int_{a}^{\,b}\!\! dz_3\, 
\widetilde{\Phi}^{''}_{\alpha \beta}(z_3)
\\
&\hspace*{-1.cm}\times \int_0^{\,\infty}\!\! 
dk'k'\, \overline{\omega}_{\alpha}(z_3, z_1, k')\, \overline{\omega}_{\beta}(z_3, z_2, k') 
\,J_0(k' r_{2}).
\notag
\end{align}
It is the second step of the calculation (Hankel transformation with respect to the external 
coordinate $r_2$) which presents difficulties. 
Applying the integral operator $2\pi\!\int_0^{\infty} dr_2\,r_2 J_0(k r_2)$ to \eqref{tricky} yields the 
Hankel transformation 
\begin{align}\label{tricky_transform}
&\!\overline{c}^{\,B}_{\alpha \beta}(z_1,z_2,k)\!=\!
\frac{1}{4\pi R^2}\!\int_{a}^{\,b} \!\!\!dz_3\, 
\widetilde{\Phi}^{''}_{\alpha\beta}(z_3)
\!\!\int_0^{\,\infty} \!\!\!dk'k'\, \overline{\omega}_{\alpha}(z_3, z_1, k') 
\notag
\\
&\hspace*{-0.2cm}\times \overline{\omega}_{\beta}(z_3, z_2, k') 
\left[ 
2\pi\!\!\int_{0}^{\,\infty} dr_2\,r_2 J_0(k r_2) \left(-r_2^2 J_0(k' r_{2})\right)
\right]\!.
\end{align}
To make progress we must evaluate the integral in square brackets; the Hankel transform of 
$-r_2^2 J_0(k' r_{2})$. 
Given a test function, $f(r)$, which vanishes sufficiently rapidly as $r\!\rightarrow\!\infty$, 
it can be shown that the Hankel transform of $-r^2 f(r)$ is given by
\begin{align}
\overline{-r^2 f}(r) &= \frac{d^2 \overline{f}(k)}{dk^2} + \frac{1}{k} \frac{d \overline{f}(k)}{dk}.  
\end{align}
Setting $f(r)=J_0(k' r)$ and using the result \eqref{orthogonality prop Hankel transf} yields
\begin{align}
\overline{-r_2^2 J_0}(k' r_{2}) = (2 \pi)^2 \left(\delta''(\vec{k}-\vec{k'}) + \frac{1}{k} \delta'(\vec{k}-\vec{k'}) \right),
\end{align}
where the prime(s) on the $\delta$-functions express the first(second) derivative with respect to $k$. 
Using the known properties of delta-function derivatives the expression \eqref{tricky_transform} thus becomes
\begin{align}\label{almost_done1}
\overline{c}^{\,B}_{\alpha \beta}(z_1,z_2,k)&=
\frac{2 \pi^2}{(4 \pi R)^{\delta_{1 |\alpha|}+\delta_{1 |\beta|}}} 
\int_{a}^{\,b} \!\!dz_3\, 
\widetilde{\Phi}^{''}_{\alpha\beta}(z_3)
\\
&\times 
\left(\frac{\partial^2}{\partial k^2} + 
\frac{1}{k} \frac{\partial}{\partial k} \right) \Big(J_0\left(k R_{13}\right) J_0\left(k R_{23}\right)\Big), 
\notag
\end{align}
where we have used the explicit expressions for the transformed weight functions 
\eqref{wH} to introduce the Bessel functions. 
The derivatives of the Bessel function product yield the following expression
\begin{align}\label{almost_done2}
&\left(\frac{\partial^2}{\partial k^2} + 
\frac{1}{k} \frac{\partial}{\partial k} \right) \Big(J_0\left(k R_{13}\right) J_0\left(k R_{23}\right)\Big) =
\\ 
&  -\frac{1}{2}\left(R_{13}^2 + R_{23}^2\right) J_0\left(k R_{13}\right) J_0\left(k R_{23}\right) 
\notag\\
&+ 2 R_{13} R_{23} J_1\left(k R_{13}\right) J_1\left(k R_{23}\right) 
\notag\\ 
&  + \frac{1}{2}R_{13}^2 J_2\left(k R_{13}\right) J_0\left(k R_{23}\right)
+ \frac{1}{2}R_{23}^2 J_0\left(k R_{13}\right) J_2\left(k R_{23}\right) 
\notag\\ 
&  - \frac{1}{k} \Bigl(R_{13} J_1\left(k R_{13}\right) J_0\left(k R_{23}\right) 
+ R_{23} J_0\left(k R_{13}\right) J_1\left(k R_{23}\right)\Bigr).
\notag
\end{align}
Putting the results \eqref{almost_done1} and \eqref{almost_done2} together with 
\eqref{class3_aim} and \eqref{partial_result} yields the final result for the Hankel transform 
of the class 3 contributions with $\alpha,\beta\in \{{\bf 1},{\bf 2}\}$
\begin{multline}\label{final}
\overline{c}_{\alpha \beta}(z_1, z_2, k) = \left(1-\frac{z_{12}^2}{2 R^2}\right) 
\mathcal A_{\alpha \beta}(z_1, z_2, k) \\
+ \frac{2 \pi^2}{(4 \pi R)^{\delta_{1 |\alpha|}+\delta_{1 |\beta|}}} 
{\mathcal B}_{\alpha \beta}(z_1, z_2, k),
\end{multline}
for $|z_{12}|\le 2R$ and zero otherwise, where 
%
\begin{align*}
&{\mathcal B}_{\alpha \beta}(z_1, z_2, k) = \int_{a}^{\,b} \!\!dz_3\, \widetilde{\Phi}^{''}_{\alpha \beta}(z_3)\\ 
&\times\bigg\{ -\frac{R_{13}^2 + R_{23}^2}{2} J_0\left(k R_{13}\right) J_0\left(k R_{23}\right)
\\ 
&+ 2 R_{13} R_{23} J_1\left(k R_{13}\right) J_1\left(k R_{23}\right)
+ \frac{R_{13}^2}{2} J_2\left(k R_{13}\right) J_0\left(k R_{23}\right)
\\
&+ \frac{R_{23}^2}{2} J_0\left(k R_{13}\right) J_2\left(k R_{23}\right)
- \frac{1}{k} \Bigl(R_{13} J_1\left(k R_{13}\right) J_0\left(k R_{23}\right)
\\ 
&+ R_{23} J_0\left(k R_{13}\right) J_1\left(k R_{23}\right)\Bigr)\bigg\}.
\end{align*}
To summarize, equations \eqref{cbar_sum}, \eqref{class1}, \eqref{class2A}, \eqref{class2B}, 
and \eqref{final} provide the Hankel transform of the FMT pair direct correlation function 
as an explicit functional of the one-dimensional planar density profile. 
Given $\overline{c}^{(2)}_{\rm hs}$ the total correlation function, 
$\overline{h}_{\rm hs}$, can be calculated using the OZ relation \eqref{oz_planar}. 
The pair correlations in real-space can then be obtained via 
(numerical) Hankel back-transformation using \eqref{hankel_back}.


\subsection*{FMT in spherical geometry}

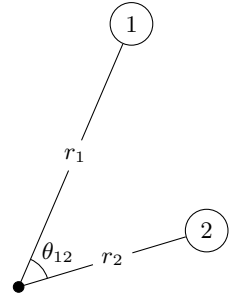
\begin{figure}
\begin{minipage}[t]{0.4\textwidth}
\hspace*{-0.5cm}
\begin{tikzpicture}
\coordinate (origine) at (0,0);
\filldraw (origine) circle (2pt);
\node[draw, circle] (particle 1) at (1.5,3.5) {1};
\node[draw, circle] (particle 2) at (2.5,0.75) {2};
\node (r1 label) at (0.75,1.75) {$r_1$};
\node (r2 label) at (1.25,0.375) {$r_2$};
\draw[-, >=latex] (origine) -- (r1 label);
\draw[-, >=latex] (r1 label) -- (particle 1);
\draw[-, >=latex] (origine) -- (r2 label);
\draw[-, >=latex] (r2 label) -- (particle 2);
\pic [draw, -, angle radius=4mm, angle eccentricity=1.75, "$\theta_{12}$"] {angle = particle 2--origine--particle 1};
\end{tikzpicture}
\end{minipage}
\hspace*{2cm}
\caption{Sketch of the spherical geometry. 
}
\label{spherical_sketch}
\end{figure}

When the external field has spherical symmetry the density only varies as a function 
of the distance from the origin. The inhomogeneous 
pair correlations thus require 
as input the two radial distances, $r_1$ and $r_2$, and the cosine 
of the angle between them, $x_{12}\!=\!\cos(\,\theta_{12})$ (see Fig.\ref{spherical_sketch}).
In planar geometry the transformed 
weight functions \eqref{wH} depend only on the separation $z_{12}$, whereas the analogous 
expressions in spherical geometry depend on both arguments $r_1$ and $r_2$. 
In addition, the transformed weight functions in spherical geometry change their 
functional forms whenever one (or both) of these arguments approaches the origin to within 
a distance $R$. 
With the intention of sparing the reader technical overload, the results to be presented below will 
be restricted to cases satisfying both $r_1\!\ge\! R$ and $r_2\!\ge\! R$, 
If other situations arise, as would be the case for e.g.~hard-spheres confined to a spherical cavity, 
then the methods to be discussed below could be easily generalized.  
In the following we will consider only test-particle calculations, for which the density, $\rho(r)$, 
is zero for $r\!<\!R$.  

The Legendre transform  of the pair direct correlation function is given by 
\begin{align}\label{legendre}
\hat{c}^{\,(2)}_{\rm hs}(r_1,r_2,n) 
= 
\frac{2n+1}{2} \int_{-1}^{1} \!dx_{12}\, P_n(x_{12})c^{(2)}_{\rm hs}(r_1,r_2,x_{12}),
\end{align}
where $P_n(x_{12})$ is a Legendre polynomial. 
The back-transform is given by 
\begin{align}\label{legendre_back}
c^{(2)}_{\rm hs}(r_1,r_2,x_{12}) 
= 
\sum_{n=0}^{\infty} P_n(x_{12})\,\hat{c}^{\,(2)}_{\rm hs}(r_1,r_2,n).
\end{align}
%
Legendre transform of the OZ equation \eqref{oz} simplifies the three-dimensional 
integral and yields the following equation 
for the transforms (see Appendix \ref{Appendix_OZspherical})
\begin{multline}\label{oz_spherical}
\hat{h}^{}_{\rm hs}(r_1,r_2,n) = \hat{c}^{\,(2)}_{\rm hs}(r_1,r_2,n) 
\\
+ \frac{4\pi}{2n+1} \int_{0}^{\infty} \!dr_3 \; r_3^2\,  
\hat{h}^{}_{\rm hs}(r_1,r_3,n)\,\rho(r_3)\,\hat{c}^{\,(2)}_{\rm hs}(r_3,r_2,n).
\end{multline}
%
%
The Legendre transform of the two-body direct correlation function \eqref{c_ros_fe} can be expressed 
as the following sum
\begin{equation}\label{chat_sum}
\hat{c}^{(2)}_{\rm hs}(r_1, r_2, n) = -\sum_{\alpha\beta} \hat{c}_{\alpha \beta}(r_1, r_2, n),
\end{equation}
where the terms in \eqref{chat_sum} are constructed using the Legendre transformed scalar 
weight functions
%
%
%
\begin{align}
\hat{\omega}_{3}(r_1, r_2, n)&\!=\!
\begin{cases}
      \frac{1}{2}\! \left(1-x_{12}\right)\!\Theta_{12} &\!\! n\!=\!0,\\
      \frac{3}{4}\! \left(1-x_{12}^2\right)\!\Theta_{12} &\!\! n\!=\!1,\\
      \frac{2n+1}{2n}\! \big(x_{12} P_n(x_{12})\!-\!P_{n+1}(x_{12})\big) \Theta_{12} &\!\! n\!\ge\! 2,
    \end{cases} 
\notag\\
\hat{\omega}_{2}(r_1, r_2, n) &\!=\! \frac{2n+1}{2}\frac{R}{r_1 r_2}P_n(x_{12}) \Theta_{12}, 
\notag\\
\hat{\omega}_{1}(r_1, r_2, n) &\!=\! \frac{\hat{\omega}_{2}(r_1, r_2, n)}{4\pi R}, 
\notag\\
\hat{\omega}_{0}(r_1, r_2, n) &\!=\! \frac{\hat{\omega}_{2}(r_1, r_2, n)}{4\pi R^2}, 
\end{align}
where $x_{12}=(r_1^2+r_2^2-R^2)/(2 r_1 r_2)$, 
$\Theta_{12}\!\equiv\!\Theta\left(R-|r_{12}|\right)$ is the Heaviside step function
and $r_{12}=r_1-r_2$.

In analogy with our treatment of planar geometry we consider separately the three classes of terms 
contributing to the sum \eqref{chat_sum}.
For terms belonging to {\bf class 1} the steps involved in transforming 
$c_{\alpha \beta}$ are identical to those required to transform the OZ equation 
(see Appendix \ref{Appendix_OZspherical}).  
For $\alpha,\beta \in \{0, 1, 2, 3\}$ this yields
\begin{align}\label{class1_sph}
\hat{c}_{\alpha \beta}(r_1, r_2, n) &= \frac{4\pi}{2n+1}\!\int_{d}^{e} \!\!dr_3 \; r_3^2 \\
&\quad \times \hat{\omega}_{\alpha}(r_3, r_1, n) \Phi^{''}_{\alpha \beta}(r_3) 
\,\hat{\omega}_{\beta}(r_3, r_2, n),
\notag
\end{align}
for $|r_{12}|\le 2R$ and zero otherwise. For the restricted ranges of $r_1$ and $r_2$ under consideration 
the integration limits are  
$d=\max(r_1,r_2)\!-\!R$ and $e=\min(r_1,r_2)\!+\!R$.

For mixed terms with $\alpha \in \{0, 1, 2, 3\}$ and $\beta \in \{{\bold 1}, {\bold 2}\}$, 
we consider the following scalar product
\begin{align*} 
\Phi^{''}_{\alpha \beta}(r_3)\cdot
\omega_{\beta}(\rv_{32})
=
|\Phi^{''}_{\alpha \beta}(r_3)|\,
\omega_{|\beta|}(\rv_{32})
\,\unit_{r_3}\cdot
\unit^{\,\rm shell}_{32},
\end{align*}
where $\unit_{r_3}\!\cdot\,\unit^{\,\rm shell}_{32}=(R^2+r_3^2-r_2^2)/(2 r_3 R)$.  
This result depends only on $r_3$ and $r_2$ and so its presence in the integrand does not 
interfere with the Legendre transformation and the standard procedure 
given in Appendix \ref{Appendix_OZspherical}
can be applied without modification. 
We thus obtain the following expression
\begin{align}\label{class2A_sph}
\hspace*{-0.15cm}\hat{c}_{\alpha \beta}(r_1, r_2, n) =&
\frac{4\pi}{2n+1}\int_{d}^{e} \!\!dr_3\, r_3^2\,
\hat{\omega}_{\alpha}(r_3, r_1, n)
\,|\Phi^{''}_{\alpha \beta}(r_3)|
\notag\\
&\times \left(\frac{R^2+r_3^2-r_2^2}{2 r_3 R}\right)\hat{\omega}_{|\beta|}(r_3, r_2, n),
\end{align}
for $|r_{12}|\le 2R$ and zero otherwise.
Similarly, when the rank of the indices is exchanged, 
$\alpha \in \{{\bold 1}, {\bold 2}\}$ and $\beta \in \{0, 1, 2, 3\}$, we find
\begin{align}\label{class2B_sph}
\hspace*{-0.25cm}\hat{c}_{\alpha \beta}(r_1, r_2, n) =&
\frac{4\pi}{2n+1}\int_{d}^{e} \!dr_3\; r_3^2\,
\hat{\omega}_{|\alpha|}(r_3, r_1, n)
\\
&\times \left(\frac{R^2+r_3^2-r_1^2}{2 r_3 R}\right)|\Phi^{''}_{\alpha \beta}(r_3)|\;\hat{\omega}_{\beta}(r_3, r_2, n),
\notag
\end{align}
for $|r_{12}|\le 2R$ and zero otherwise. 
\begin{figure}[!t]
\begin{minipage}[t]{0.4\textwidth}
\hspace*{-0.5cm}
\begin{tikzpicture}
\coordinate (origine) at (0,0);
\coordinate (particle 1) at (1.25,4);
\coordinate (particle 2) at (3,2.75);
\coordinate (particle 3) at (1.5,2.5);
\coordinate (label R 1) at (1.375,3.25);
\coordinate (label R 2) at (2.35,2.625);
\coordinate (label r1) at (0.625,2);
\coordinate (label r2) at (1.5,1.375);
\coordinate (label r3) at (0.9,1.5);
\draw (origine) node {$\bullet$};
\draw (particle 1) node {$\bullet$};
\draw (particle 2) node {$\bullet$};
\draw (particle 3) node {$\bullet$};
\draw[-, >=latex] (particle 1) -- (particle 2);
\draw[dashed] (particle 1) -- (particle 3) -- (particle 2);
\draw[->, >=latex] (origine) -- (particle 1);
\draw[->, >=latex] (origine) -- (particle 2);
\draw[->, >=latex] (origine) -- (particle 3);
\draw (origine) node[below] {$O$};
\draw (particle 1) node[above right] {$1$};
\draw (particle 2) node[above right] {$2$};
\draw (particle 3) node[below right] {$3$};
\draw (label r1) node[left] {$\vec{r}_1$};
\draw (label r2) node[below right] {$\vec{r}_2$};
\draw (label r3) node[right] {$\vec{r}_3$};
\draw (label R 1) node[below left] {$R$};
\draw (label R 2) node[below left] {$R$};
\pic [draw, -, angle radius=3mm, angle eccentricity=1.5, "$\alpha$"] {angle = particle 2--particle 3--particle 1};
\pic [draw, -, angle radius=5mm, angle eccentricity=1.4, "$\theta_2$"] {angle = particle 2--origine--particle 1};
\end{tikzpicture}
\end{minipage}
\hspace*{2cm}
\caption{Geometrical sketch for evaluation of the scalar product given in \eqref{cosalpha}. 
We orient our spherical coordinate system such that the $z$-axis lies along 
$\vec{r}_1$.
} 
\label{cosalpha_sketch}
\end{figure}
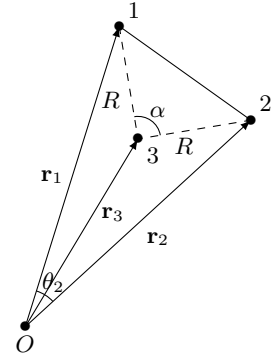	

As in the planar case, the terms in {\bf class 3} are more difficult to deal with and 
we must consider the following scalar product
%
\begin{align}\label{scalarproduct_sph}
&\omega_{\alpha}(\rv_{31})\cdot
\Phi^{''}_{\alpha\beta}(r_3)\cdot
\omega_{\beta}(\rv_{32})=
\\ 
&
\widetilde{\Phi}^{''}_{\alpha\beta}(r_3)
\,\omega_{|\alpha|}(\rv_{31}) \,\omega_{|\beta|}(\rv_{32})\;
\unit^{\,\rm shell}_{31}
\cdot
\unit^{\,\rm shell}_{32}.
\notag
\end{align}
In Fig.\ref{cosalpha_sketch} we sketch the three vectors $\rv_1$, $\rv_2$ and $\rv_3$. 
The values of $\rv_3$ which yield a nonzero contribution to 
\eqref{scalarproduct_sph} lie on the intersection circle of delta-shells centered at $\rv_1$ and $\rv_2$. 
If we choose $\rv_1$ along the $z$-axis of our spherical coordinate system, then for $\rv_3$ anywhere 
on the intersection circle we find
\begin{align}\label{cosalpha}
\hspace*{-0.16cm}\unit^{\,\rm shell}_{31}
\cdot
\unit^{\,\rm shell}_{32}
\equiv \cos(\alpha)
=1-\frac{r_1^2+r_2^2}{2 R^2}+\frac{r_1 r_2}{R^2} P_1(x_2), 
\end{align}
where the angle $\alpha$ is defined in Fig.\ref{cosalpha_sketch} and $x_2=\cos(\theta_2)$. 
We thus seek to evaluate the Legendre transform of  
\begin{align}\label{class3_aim_sph}
c_{\alpha\beta}(r_1,r_2,x_2) = c_{\alpha\beta}^{\,D}(r_1,r_2,x_2) + c_{\alpha\beta}^{\,E}(r_1,r_2,x_2),
\end{align} 
where the two contributions are given by 
\begin{align*}
c_{\alpha\beta}^{\,D}(r_1,r_2,x_2)=
\!\!\int &d\rv_3\, \widetilde{\Phi}^{''}_{\alpha\beta}(r_3)\\
&\times \omega_{|\alpha|}(\rv_{31})\, 
\omega_{|\beta|}(\rv_{32})
\!\!\left(
\!1\!-\!\frac{r_1^2+r_2^2}{2 R^2}
\!\right)\!,
\notag\\
c_{\alpha\beta}^{\,E}(r_1,r_2,x_2)=
\!\!\int &d\rv_3\, \widetilde{\Phi}^{''}_{\alpha\beta}(r_3)\\
&\times \omega_{|\alpha|}(\rv_{31})\, 
\omega_{|\beta|}(\rv_{32})
\!\left(
\frac{r_1 r_2}{R^2} P_1(x_2)
\right)\!.
\end{align*}
\!\!The factor $1-(r_1^2+r_2^2)/(2 R^2)$ is independent of $x_2$ and Legendre transformation 
thus proceeds in the same way as for the OZ equation (see Appendix \ref{Appendix_OZspherical}). 
This yields
\begin{align}\label{partial_result_sph}
\hat{c}^{\,D}_{\alpha \beta}(r_1, r_2, n) = \left(\!1\!-\!\frac{r_1^2+r_2^2}{2 R^2}\right) 
\mathcal D_{\alpha \beta}(r_1, r_2, n),
\end{align}
for $|r_{12}|\le 2R$ and zero otherwise, where the function ${\mathcal D}_{\alpha \beta}$ is given by 
\begin{align*}
{\mathcal D}_{\alpha \beta}(r_1, r_2, n)\!&=\! \frac{4\pi}{2n+1}\int_{0}^{\infty}\!\!\!dr_3\, r_3^2 \\ 
&\quad\times \widetilde{\Phi}^{''}_{\alpha \beta}(r_3) \hat{\omega}_{|\alpha|}(r_3, r_1, n) 
\,\hat{\omega}_{|\beta|}(r_3, r_2, n).
\end{align*}
Legendre transform of $c_{\alpha\beta}^{E}$ is more difficult due to the presence of the factor 
$P_1(x_2)$. 
If we follow the procedure of Appendix \ref{Appendix_OZspherical} we find that the first step of the calculation 
can be carried through easily to obtain 
\begin{align}\label{tricky_sph}
&c^{\,E}_{\alpha \beta}(r_1, r_2, x_2)=
2\pi \int_{0}^{\infty} \!dr_3\; r_3^2\,
\widetilde{\Phi}^{''}_{\alpha \beta}(r_3)
\\
&\!\times \sum_{i=0}^{\infty} \frac{2}{2i+1} \hat{\omega}_{|\alpha|}(r_3, r_1, i) 
\,\hat{\omega}_{|\beta|}(r_3, r_2, i)
\frac{r_1 r_2}{R^2} P_1(x_2) P_i(x_2).
\notag
\end{align}
Applying the operator $\frac{2n+1}{2}\!\int_{-1}^{\,1} dx_2\,P_n(x_2)$ to \eqref{tricky_sph} then yields the 
Legendre transformation 
\begin{align}\label{tricky_transform_sph}
\hat{c}^{\,E}_{\alpha \beta}(r_1,r_2,n)\!&=\!
2\pi \frac{2n+1}{2}\!\int_{0}^{\infty} \!\!\!dr_3\, r_3^2\, 
\widetilde{\Phi}^{''}_{\alpha \beta}(r_3)
\notag \\
&\times \sum_{i=0}^{\infty} \frac{2}{2i+1} \hat{\omega}_{|\alpha|}(r_3, r_1, i) 
\,\hat{\omega}_{|\beta|}(r_3, r_2, i)
\notag \\
&\times \frac{r_1 r_2}{R^2}
\!\!\int_{-1}^{\,1} \!\!\!dx_2\, P_1(x_2) P_n(x_2) P_i(x_2).
\end{align}
The extra complication here is caused by the integral of a triple product of Legendre polynomials. 
Fortunately, this integration has been well-studied in the context of quantum mechanics and can be 
reexpressed using the Wigner $3j$ notation (see e.g.~\cite{sakurai}) 
\begin{align}
\int_{-1}^{\,1} \!\!\!dx_2\, P_l(x_2) P_n(x_2) P_i(x_2) = 
2{\begin{pmatrix}
l & n & i \\ 
0 & 0 & 0
\end{pmatrix}}^2.
\end{align}
For the case of interest here, $l\!=\!1$, there are only two terms in the sum over $i$ appearing in 
\eqref{tricky_transform_sph}. 
This leads to the result
\begin{align}\label{almost_done2_sph}
\hat{c}^{\,E}_{\alpha \beta}(r_1, r_2, n) = \frac{r_1 r_2}{R^2} 
{\mathcal E}_{\alpha \beta}(r_1, r_2, n),
\end{align}
for $|r_{12}|\le 2R$ and zero otherwise, where 
\begin{align}
&{\mathcal E}_{\alpha \beta}(r_1, r_2, n) = \frac{4 \pi}{2n+1}\!\int_{0}^{\infty} \!\!\!dr_3\, r_3^2\, 
\widetilde{\Phi}^{''}_{\alpha \beta}(r_3) 
\\
&\times
\bigg( \frac{n (2n+1)}{(2n-1)^2} \hat{\omega}_{|\alpha|}(r_3,r_1,n\!-\!1) \, \hat{\omega}_{|\beta|}(r_3,r_2,n\!-\!1)
\bigg. \Biggr. 
\notag\\ 
\Biggl. \bigg.
&+ \frac{(n+1)(2n+1)}{(2n+3)^2} \hat{\omega}_{|\alpha|}(r_3,r_1,n\!+\!1) \, 
\hat{\omega}_{|\beta|}(r_3,r_2,n\!+\!1)
\bigg).
\notag
\end{align} 
Putting together \eqref{class3_aim_sph}, \eqref{partial_result_sph} and \eqref{almost_done2_sph} 
yields the final result for the Legendre transform 
of the class 3 contributions with $\alpha,\beta\in \{{\bf 1},{\bf 2}\}$
\begin{multline}\label{final_sph}
\hat{c}_{\alpha \beta}(r_1, r_2, n) = \left(\!1\!-\!\frac{r_1^2+r_2^2}{2 R^2}\right) 
\mathcal D_{\alpha \beta}(r_1, r_2, n) \\
+ \frac{r_1 r_2}{R^2} 
{\mathcal E}_{\alpha \beta}(r_1, r_2, n),
\end{multline}
for $|r_{12}|\le 2R$ and zero otherwise. 
In summary, equations \eqref{chat_sum}, \eqref{class1_sph}, \eqref{class2A_sph}, \eqref{class2B_sph}, 
and \eqref{final_sph} provide the Legendre transform of the FMT pair direct correlation function, 
$\hat{c}^{(2)}_{\rm hs}$, as an explicit functional of the one-dimensional spherical density profile, 
$\rho(r)$.




\subsection*{Numerical consistency checks}

The formulae presented in the previous subsections for the Hankel and Legendre transforms of the 
pair direct correlation functions are, in principle, straightforward to implement. However, when developing 
numerics to evaluate the remaining one-dimensional integrals it 
is useful to have some checks and limiting cases to help eliminate possible coding errors. 
In bulk there are two helpful benchmarks: 
(i) In the low density limit the pair direct correlation function reduces to the Mayer function \cite{Hansen06} 
for which both the Hankel and Legendre transforms are known exactly. 
(ii) At finite density the (real-space) analytic expression for the PY pair direct correlation function \cite{Hansen06} 
can be numerically Hankel/Legendre transformed using \eqref{hankel} and \eqref{legendre}, respectively. 
The result thus obtained should agree with the predictions of our analytical expressions.  
Contributions arising from class 2 terms (scalar-vector combinations) vanish in bulk and can thus only be tested by 
considering inhomogeneous density profiles.  
A useful check is the following relation between the one- and two-body direct correlation functions 
\begin{equation}\label{lmbw}
\nabla_{1} c^{(1)}(\vec{r}_1) = \int d\vec{r}_2 \, c^{(2)}(\vec{r}_1, \vec{r}_2) \, \nabla_{2} \rho(\vec{r}_2), 
\end{equation}
known as the Lovett-Mou-Buff-Wertheim sum-rule \cite{Widom}.  
\begin{figure}[!t]
\vspace*{-0.6cm}
\centering
\includegraphics[width=1\linewidth]{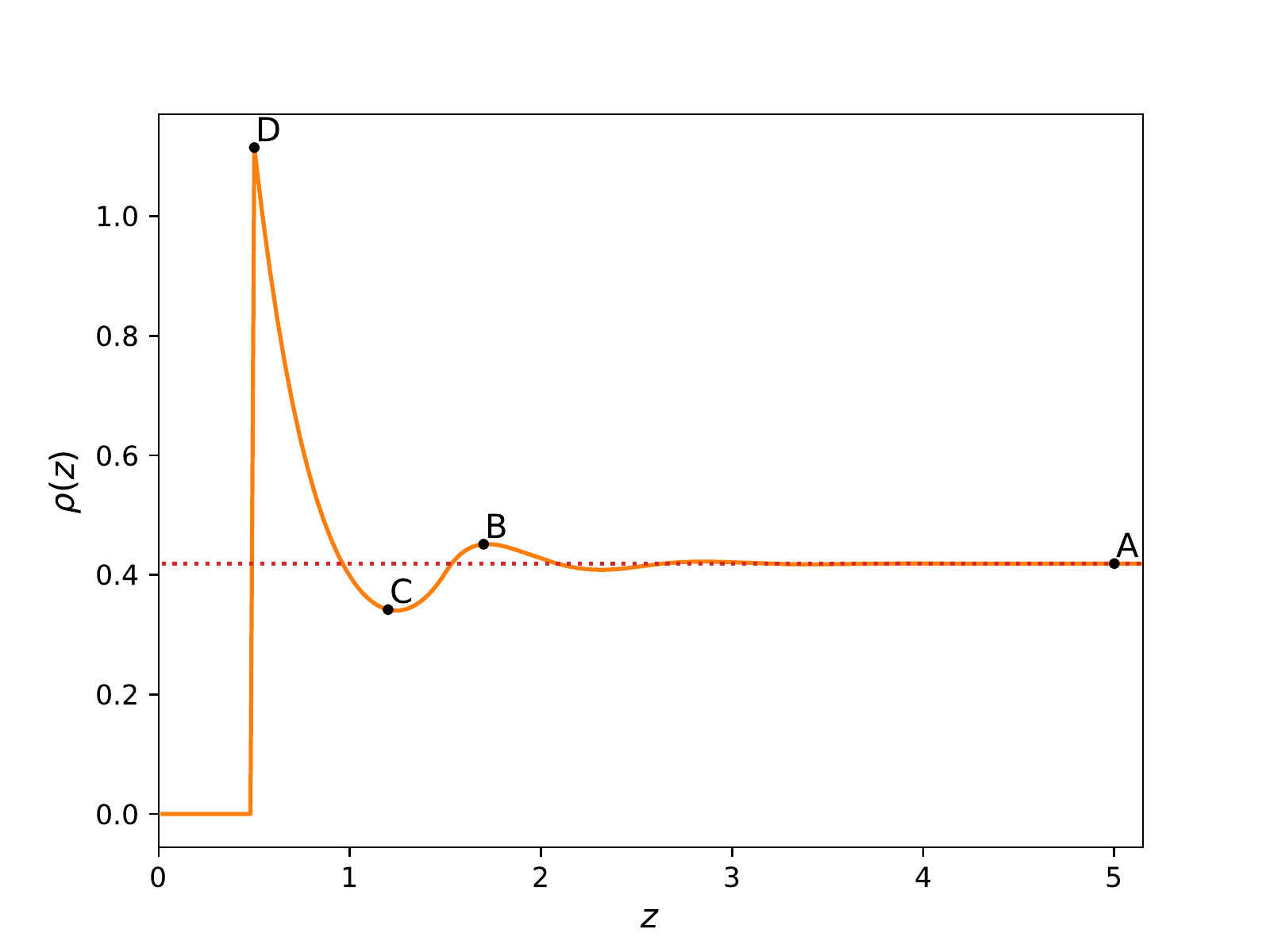}
\caption{
{\bf Hard-spheres at a hard-wall.}  
FMT density profile for $\mu=2$ (solid line) with its mid-point value indicated (horizontal dotted line). 
Points indicate the $z$-coordinates for which we show the total correlation function 
in Fig.\ref{fig: h on a wall}. 
}
\label{fig: density on a wall}
\end{figure}
In planar geometry this reduces to 
\begin{equation}
\frac{\partial c^{(1)}(z_1)}{\partial z_1} = \int_{-\infty}^{\infty} dz_2 \, \frac{\partial \rho(z_2)}{\partial z_2} \, 
\overline{c}^{(2)}(z_1, z_2, k\!=\!0), 
\end{equation}
where both $c^{(1)}$ and $c^{(2)}$ are evaluated at the equilibrium density, $\rho$. 
In spherical geometry equation \eqref{lmbw} becomes 		
\begin{equation}
\frac{\partial\,\hat{c}^{(1)}(r_1)}{\partial r_1} = 
\frac{4\pi}{3}\!\int_{0}^{\infty} \!\!dr_2 \,r_2^2\; \frac{\partial\rho(r_2)}{\partial r_2}\, \hat{c}(r_1,r_2,n\!=\!1).
\end{equation}
Finally, in planar geometry a transverse structure factor can be defined as \cite{evans79}
\begin{align}
H(z_1,k) &\equiv 1+\int dz_2 \, \overline{h}(z_1,z_2,k) \, \rho(z_2) 
\label{tsf1}
\\
&= 1+\int dz_2 \, H(z_2,k) \, \rho(z_2) \, \overline{c}^{(2)}(z_2,z_1,k), 
\label{tsf2}
\end{align}
where the second equality is an integral equation requiring iterative solution. 
The transverse structure factor is related to the local compressibility according to
\begin{align}\label{local_comp}
 H(z,k\!=\!0) = \frac{1}{\beta \rho(z)} \, \frac{\partial \rho(z)}{\partial \mu}.
\end{align} 
Satisfying equations \eqref{tsf1} and \eqref{local_comp} provides an additional check that the numerical 
solution of the OZ equation \eqref{oz_planar} for $\bar{h}_{\rm hs}$ has been performed correctly.




\section{Results for hard-spheres}\label{results_hs}

\subsection*{Planar geometry}

In Fig.\ref{fig: density on a wall} we show a FMT density profile for hard-spheres at a hard-wall.  
The chosen value of the chemical potential ($\mu\!=\!2$) corresponds to a liquid-state of intermediate bulk 
density. 
In the vicinity of the wall we observe the familiar packing oscillations which then decay rapidly into the bulk.   
The four points marked on the curve indicate the positions at which we will investigate the inhomogeneous 
two-body correlations. 
%
\begin{figure}[!t]
\centering
\includegraphics[width=1\linewidth]{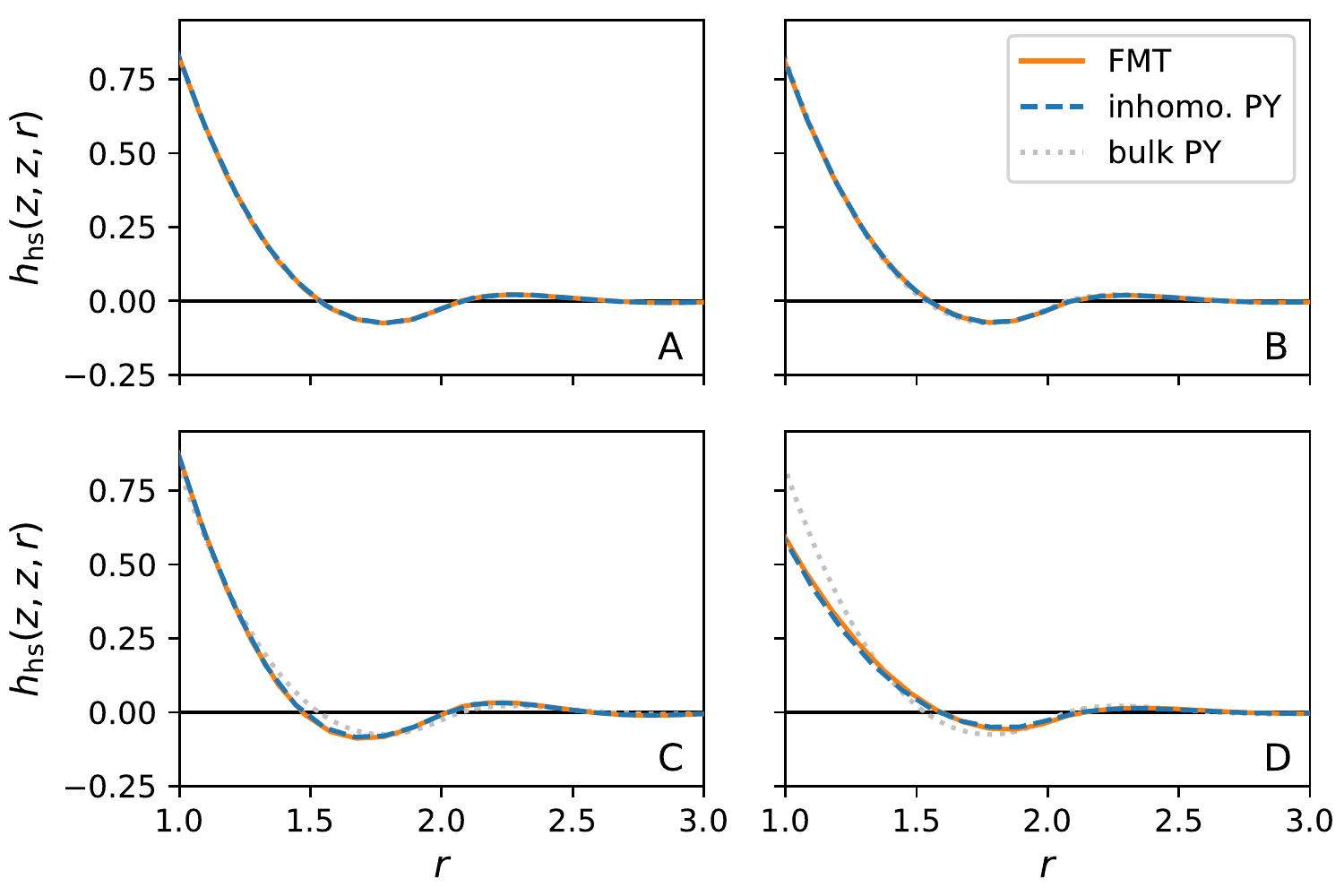}
\caption{
{\bf Hard-spheres at a hard-wall.} 
Total correlation function for equal values of the two $z$-arguments, $h_{\rm hs}(z,z,r)$, 
corresponding to the $z$-positions indicated on the density profile shown in Fig.\ref{fig: density on a wall}.  
FMT (full orange line), inhomogeneous PY (dashed blue line) and PY bulk solution (grey dotted line). 
At points A and B all three curves remain essentially identical. At points C and D there are deviations 
from bulk, but excellent agreement between FMT and PY theory.
}
\label{fig: h on a wall}
\end{figure}
When analyzing two-body correlations with planar symmetry we are faced with a function of three 
independent scalar arguments. 
This naturally presents many alternatives for graphical 
representation of the data.  
Following a quite extensive study of these various possibilities we have come to the conclusion 
that simple one-dimensional plots showing the variation of the correlation functions as a function of 
$r$ for equal values of the $z$-coordinates provides a reasonable way to compare different theories.  
Similar plots for fixed, but distinct, 
values of the $z$-coordinates were not found to offer any greater insight.

In Fig.\ref{fig: h on a wall}, we show the total correlation function for $z_1\!=\!z_2\!=\!z$ 
as a function of the cylindrical radial coordinate, $r$. 
In each panel we indicate the bulk function, 
$h_{\rm hs}(z\!\rightarrow\!\infty,z\!\rightarrow\!\infty,r)$, 
as a visual reference. 
Moving through the panels from A to D we observe increasing deviation of the inhomogeneous total 
correlation function from its bulk form. 
At all of the considered points the FMT prediction stays very close to that of the inhomogeneous 
PY theory, even when the density is strongly varying. 
This good level of agreement between the PY theory and FMT gives us some confidence in the quality 
of FMT at the two-body level, at least for these intermediate densities.
We note that the PY total correlation function of hard-spheres is unique, in the sense that it 
is generated by a strictly truncated direct correlation for $r_{12}>1$ while still satisfying the core 
condition (see \eqref{py}). 
Due to the finite range of the weight functions the FMT direct correlation function 
automatically satisfies the first of these conditions, but not the second (except in the low density 
limit). 
It thus follows that any deviation of the FMT total correlation function from the PY theory is 
a consequence of core condition violation.
The good level of agreement shown in Fig.\ref{fig: h on a wall} can be therefore taken as an indirect 
indication that the core condition is well approximated by FMT for the considered density.

\begin{figure}[!t]
\centering
\includegraphics[width=1\linewidth]{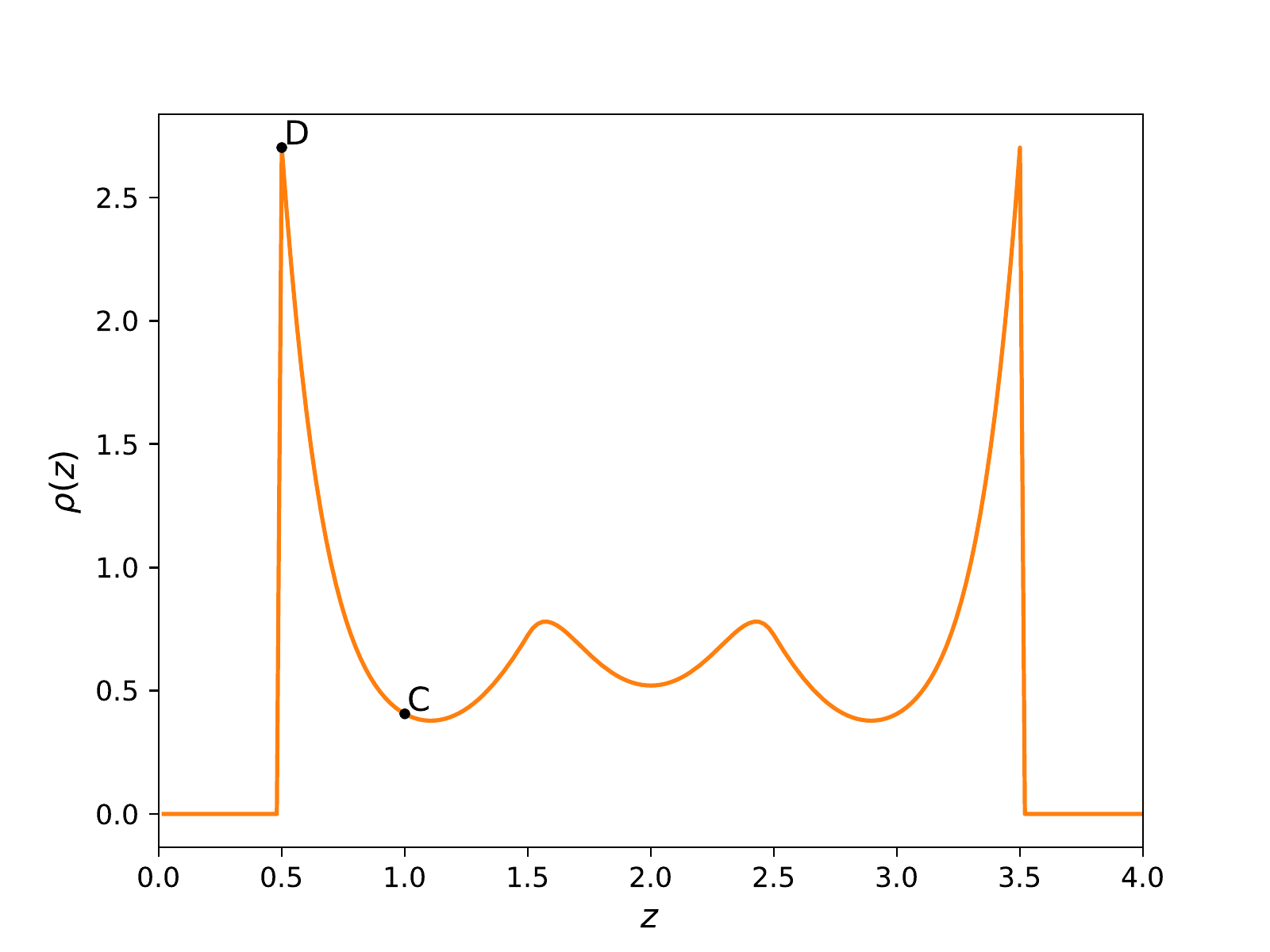}
\caption{
{\bf Confined hard-sphere system.}  
FMT density profile for $\mu=5$ (solid line). 
Points indicate the $z$-coordinates for which we show the total correlation function in 
Fig.\ref{fig: h confined syst}.
}
\label{fig: density confined syst}
\end{figure}

\begin{figure}[!t]
\centering
\includegraphics[width=1\linewidth]{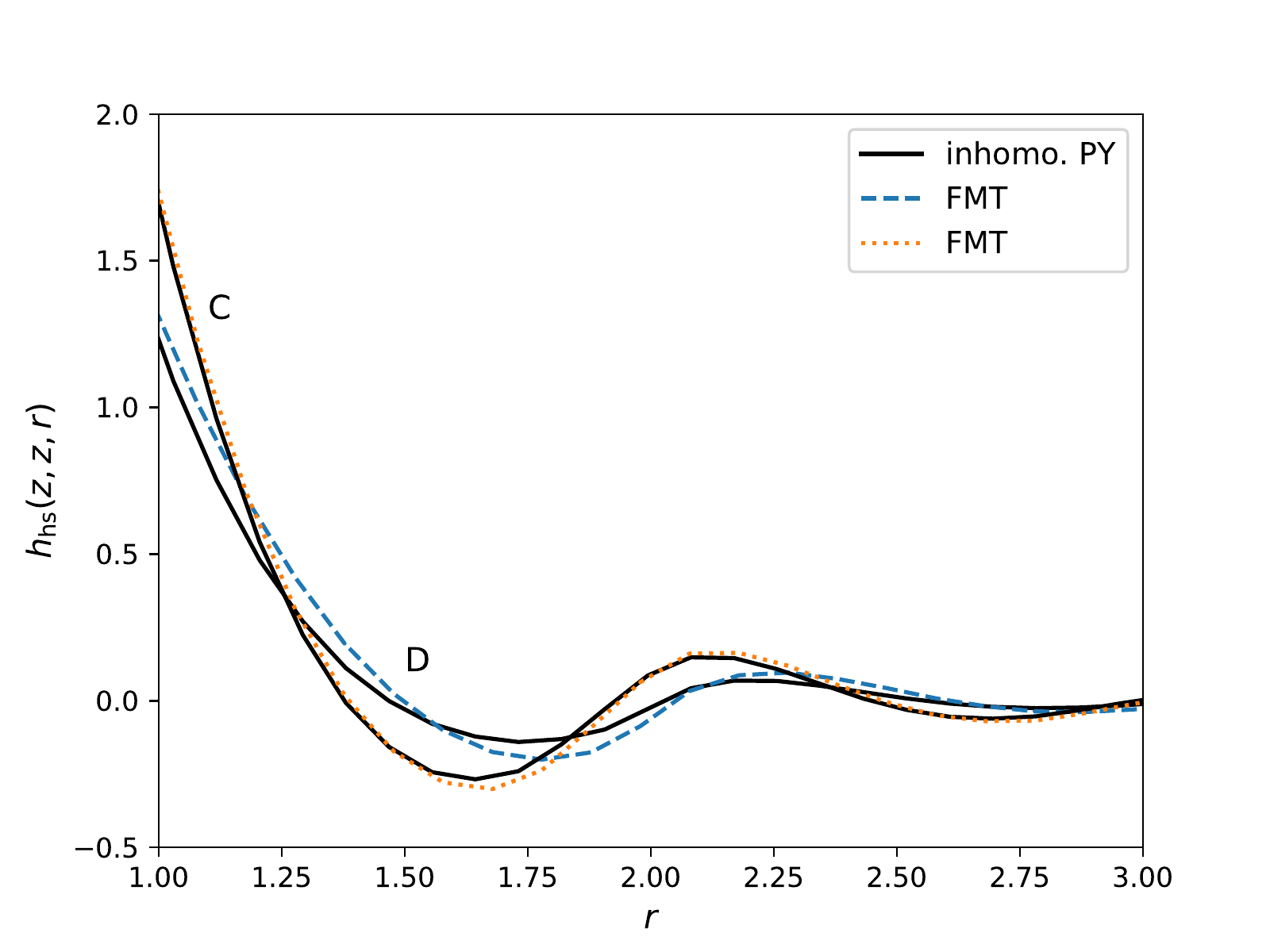}
\caption{
{\bf Confined hard-sphere system.}
Two-body total correlation function corresponding to the density profile shown in 
Fig.\ref{fig: density confined syst}. 
Inhomogeneous PY solutions 
(solid black lines) and FMT for $z=0.5$ (dashed blue line) and for $z=1$ (dotted orange line). 
We observe close agreement between FMT and PY theory at point C, but deviations between the two theories 
emerge at point D. 
}
\label{fig: h confined syst}
\end{figure}

We next consider a more demanding case: densely packed hard-spheres confined 
between two parallel hard-walls separated by four particle diameters. 
In Fig.\ref{fig: density confined syst} we show the density calculated at $\mu=5$, 
which generates a strongly inhomogeneous profile. The points label the positions 
at which we will investigate the inhomogeneous total correlation function. 
In Fig.\ref{fig: h confined syst} we show $h_{\rm hs}$ as a function of $r$ for 
$z\!=\!1$ and $z\!=\!0.5$ (the positions 
labelled C and D in Fig.\ref{fig: density confined syst}). 
At the point C, close to the first minimum of the profile, we find very close agreement 
between the PY theory and FMT, with only slight deviation at around $r\!=\!1.75$. 
At point D, corresponding to the contact peak of the profile, we find more substantial 
differences between the two approaches. The amplitude of the oscillations predicted by 
the FMT are somewhat larger than those from the PY theory, but the overall level of agreement 
remains satisfactory. For separations $r>2.75$ the predictions of PY theory and the FMT become 
very similar.

\subsection*{Spherical geometry}

As a test of our analytic FMT formulae in spherical geometry we will use the inhomogeneous total correlation 
function to calculate the three-body correlations of the bulk fluid. This can be achieved by extending 
the test-particle idea of Percus \cite{fundamentals} to the two-body level. 
If we specify the external field to represent a hard-sphere fixed at the coordinate 
origin, then the inhomogeneous correlation function 
$g^{tp}_{\rm hs}(\rv_1,\rv_2)\equiv h^{tp}_{\rm hs}(\rv_1,\rv_2)+1$ is related to the {\it bulk} triplet correlation 
function according to
\begin{eqnarray}
g^{(3)}(r_1,r_2,r_{12})=\frac{\rho^{tp}(r_1)\rho^{tp}(r_2)g^{tp}(\rv_1,\rv_2)}{\rho^2_b}, 
\label{triplet}
\end{eqnarray}
where we employ the superscript $tp$ to indicate functions calculated in the presence of a test-particle 
at the origin. 
Experience with triplet correlations has shown that direct analysis of $g^{(3)}$ is not the best choice 
when seeking to assess the quality of a given approximation. A better option is the following function
\begin{eqnarray}
\Gamma(r_1,r_2,r_{12})=\frac{\rho_b^3\, g^{(3)}\!(r_1,r_2,r_{12})}{\rho^{tp}(r_1)\,\rho^{tp}(r_2)\,\rho^{tp}(r_{12})},
\label{big_gamma}
\end{eqnarray}
which scales the triplet correlation function by the well-known Kirkwood superposition 
approximation \cite{kirkwood}. 
Deviations of $\Gamma$ from unity thus provide a sensitive measure of nontrivial contributions to 
the three-body correlations.

\begin{figure}
    \begin{minipage}[t]{1\linewidth}
    \begin{tikzpicture}
    \node[inner sep=0pt] (fig10) at (0,0)
        {\includegraphics[width=1\linewidth]{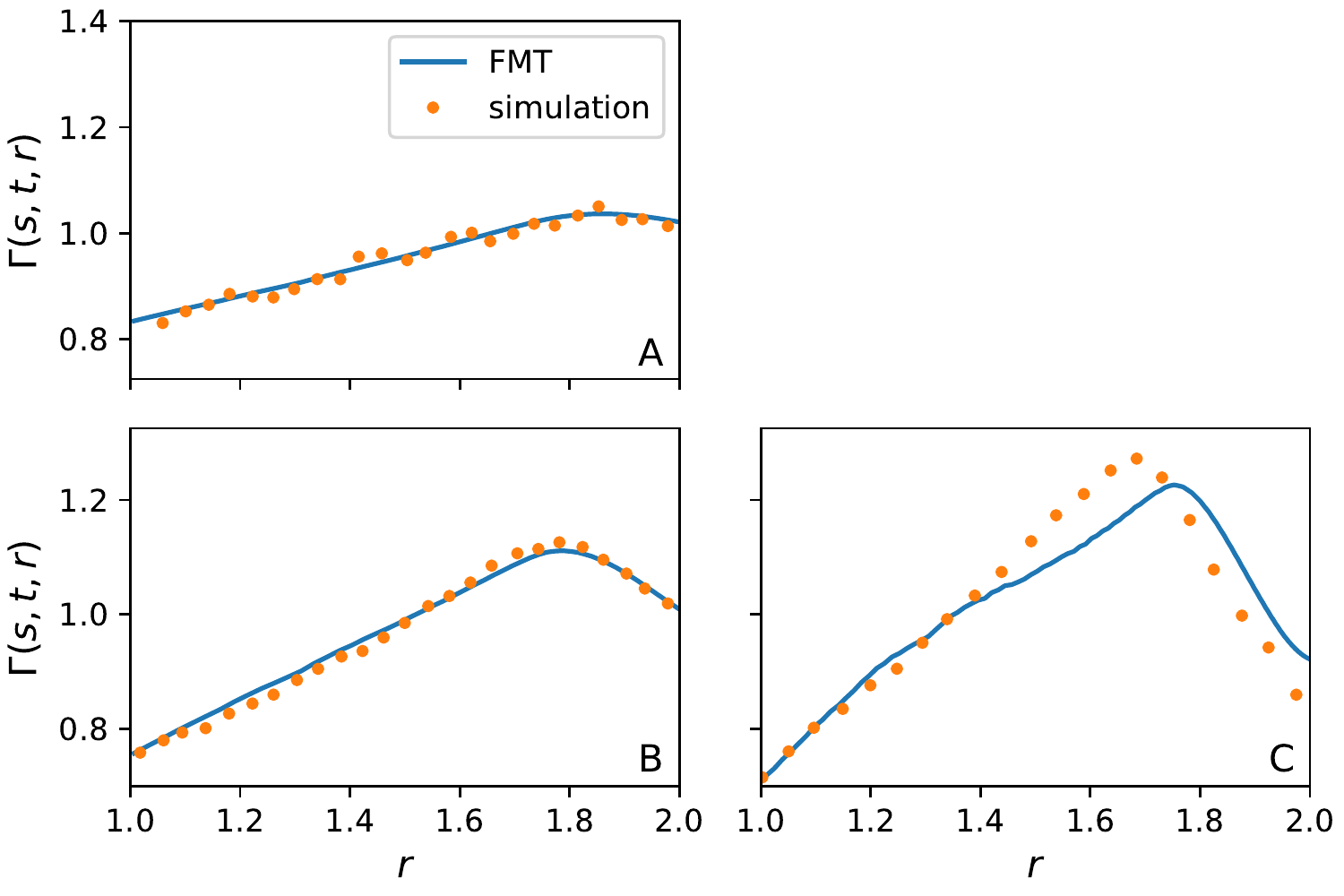}};
    \coordinate (particle 1a) at (2.2,1.425);
    \coordinate (particle 2a) at (2.2,2.13);
    \coordinate (particle 3a) at (2.86,1.675);
    \draw (particle 1a) node {$\bullet$};
    \draw (particle 2a) node {$\bullet$};
    \draw (particle 3a) node {$\bullet$};
    \draw (particle 1a) circle (10pt);
    \draw (particle 2a) circle (10pt);
    \draw (particle 3a) circle (10pt);
    \draw[densely dotted] (particle 1a) circle (20pt);
    \draw[<->, >=latex] (particle 2a) -- (particle 3a);
    \end{tikzpicture}
    \end{minipage}
    \caption{
{\bf Hard-sphere triplet correlations.} 
Comparison of FMT (lines) with simulation data (points) \cite{gubbins,kahl1994} 
for the quantity $\Gamma(1,1,r)$ at $\rho_b\!=\!0.3$ (A), $0.5$ (B) and $0.7$ (C).  
We consider rolling contact configurations for which the separation $r$ is indicated 
by an arrow in the sketch. 
}
    \label{fig: triplet 2}
    \end{figure}
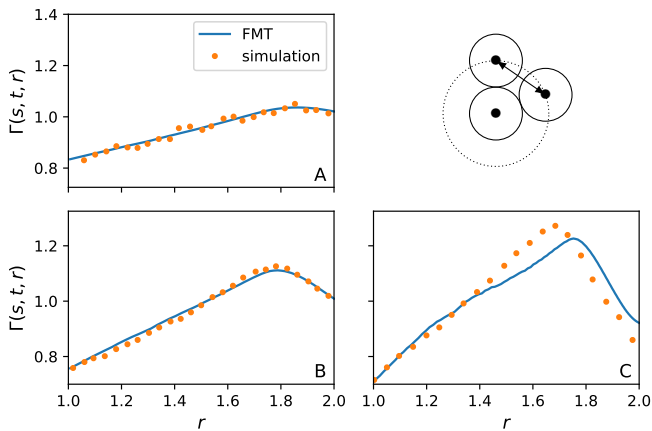

In Fig.\ref{fig: triplet 2} we show the function $\Gamma$ generated by FMT 
for `rolling contact' configurations at bulk densities $\rho_b\!=\!0.3, 0.5$ and $0.7$ (marked 
A, B and C, respectively, in the figure). 
These configurations are where the Kirwood superposition (independent probability) approximation 
is most severely tested, but are also of central importance in kinetic theories for the transport 
properties of hard-spheres (see e.g.~\cite{deschepper,leegwater}). The FMT predictions are compared with 
Monte-Carlo simulation data taken from Refs.~\cite{gubbins} and \cite{kahl1994}. 
For the two lower bulk densities considered (points A and B) we find a good level of agreement between 
FMT and simulation. 
However, at $\rho_b\!=\!0.7$ discrepancies emerge and the FMT prediction for the amplitude and position 
of the peak is less accurate. 
This suggests that we are approaching the limit at which the FMT two-body correlations can be considered 
reliable. 

\begin{figure}
\vspace*{-0.6cm}
\begin{minipage}[t]{1\linewidth}
\begin{tikzpicture}
\node[inner sep=0pt] (fig10) at (0,0)
    {\includegraphics[width=1\linewidth]{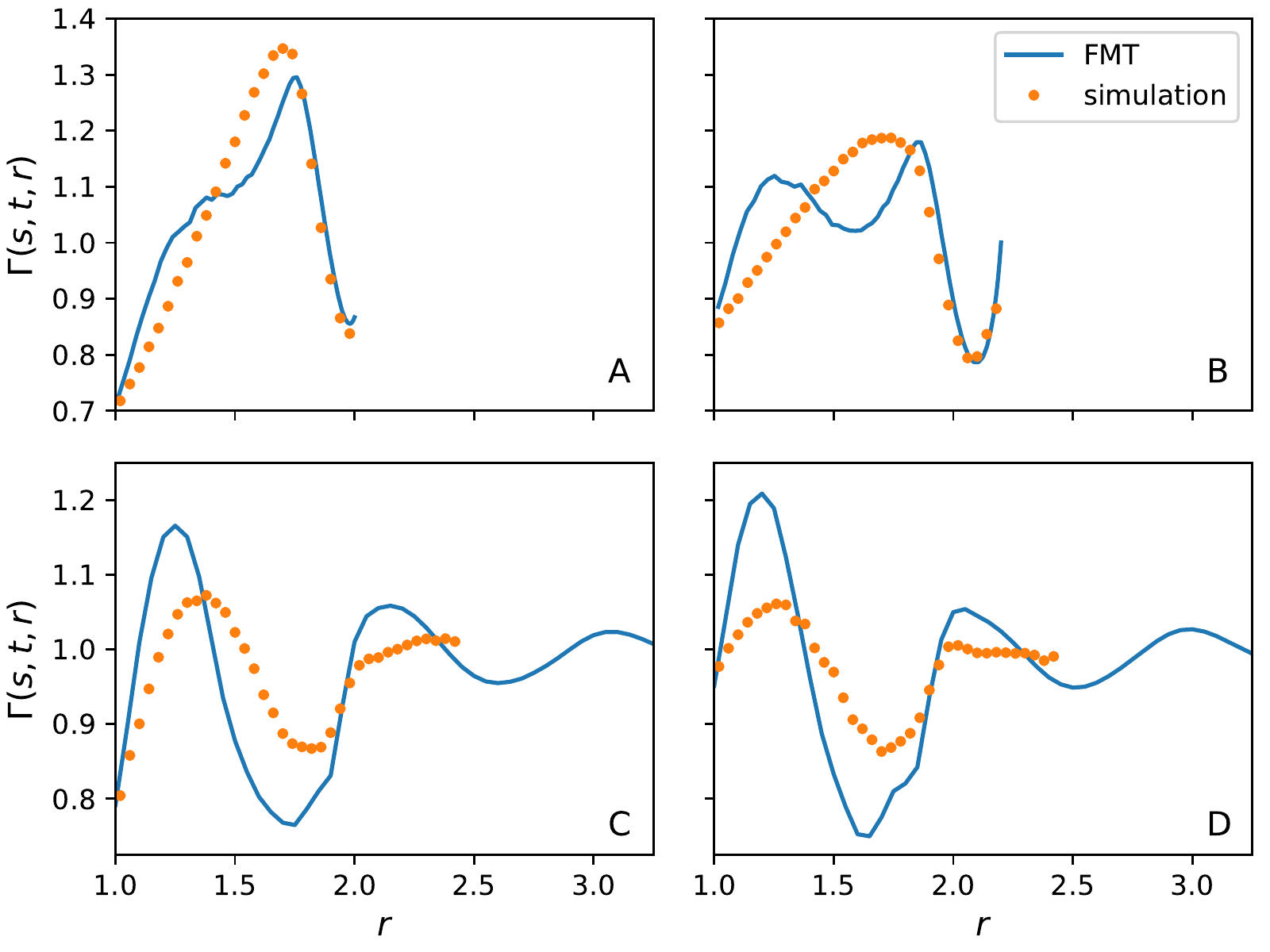}};
\scalebox{0.6}{
\coordinate (particle 1a) at (-1.5,2.8-0.15);
\coordinate (particle 2a) at (-1.5,3.505-0.15);
\coordinate (particle 3a) at (-0.84,3.05-0.15);
\draw (particle 1a) node {$\bullet$};
\draw (particle 2a) node {$\bullet$};
\draw (particle 3a) node {$\bullet$};
\draw (particle 1a) circle (10pt);
\draw (particle 2a) circle (10pt);
\draw (particle 3a) circle (10pt);
\draw[densely dotted] (particle 1a) circle (20pt);
\draw[<->, >=latex] (particle 2a) -- (particle 3a);
\coordinate (particle 1b) at (5.5-0.15,2.8-0.15);
\coordinate (particle 2b) at (5.5-0.15,3.65-0.15);
\coordinate (particle 3b) at (6.35-0.15,3.05-0.15);
\draw (particle 1b) node {$\bullet$};
\draw (particle 2b) node {$\bullet$};
\draw (particle 3b) node {$\bullet$};
\draw (particle 1b) circle (10pt);
\draw (particle 2b) circle (10pt);
\draw (particle 3b) circle (10pt);
\draw[densely dotted] (particle 1b) circle (25pt);
\draw[<->, >=latex] (particle 2b) -- (particle 3b);
\coordinate (particle 1c) at (-2,-1.25-0.05);
\coordinate (particle 2c) at (-2,-0.45-0.05);
\coordinate (particle 3c) at (-1,-0.85-0.05);
\coordinate (middle 12c) at (-2,-0.85-0.05);
\coordinate (end 12c 3c) at (0,-0.85-0.05);
\coordinate (start 12c 3c) at (-1.41,-0.85-0.05);
\draw (particle 1c) node {$\bullet$};
\draw (particle 2c) node {$\bullet$};
\draw (particle 3c) node {$\bullet$};
\draw (particle 1c) circle (10pt);
\draw (particle 2c) circle (10pt);
\draw (particle 3c) circle (10pt);
\draw[|->, >=latex, densely dotted] (start 12c 3c) -- (end 12c 3c);
\draw[<->, >=latex] (particle 1c) -- (particle 3c);
\draw[<->, >=latex] (particle 2c) -- (particle 3c);
\coordinate (particle 1d) at (4.95-0.15,-1.35-0.05);
\coordinate (particle 2d) at (4.95-0.15,-0.35-0.05);
\coordinate (particle 3d) at (5.95-0.15,-0.85-0.05);
\coordinate (middle 12d) at (4.95-0.15,-0.85-0.05);
\coordinate (end 12d 3d) at (6.95-0.15,-0.85-0.05);
\coordinate (start 12d 3d) at (5.54-0.15,-0.85-0.05);
\draw (particle 1d) node {$\bullet$};
\draw (particle 2d) node {$\bullet$};
\draw (particle 3d) node {$\bullet$};
\draw (particle 1d) circle (10pt);
\draw (particle 2d) circle (10pt);
\draw (particle 3d) circle (10pt);
\draw[|->, >=latex, densely dotted] (start 12d 3d) -- (end 12d 3d);
\draw[<->, >=latex] (particle 1d) -- (particle 3d);
\draw[<->, >=latex] (particle 2d) -- (particle 3d);
}
\end{tikzpicture}
\end{minipage}
\caption{\textbf{Hard-sphere triplet correlations.} 
Comparison of the FMT (lines) with simulation data (points) \cite{gubbins} for the quantity $\Gamma(s,t,r)$ at 
$\rho_b=0.8$. Configurations are sketched in each figure. 
The separation $r$ is indicated by the bold arrow. 
A and B are rolling geometries at 
$s=t=1.0$ and $s=t=1.1$, respectively. C and D are isosceles triangle configurations with $s=r$ 
and the base length of the triangle fixed at $t=1.1$ and $1.3$, respectively. 
}
\label{fig: triplet}
\end{figure}
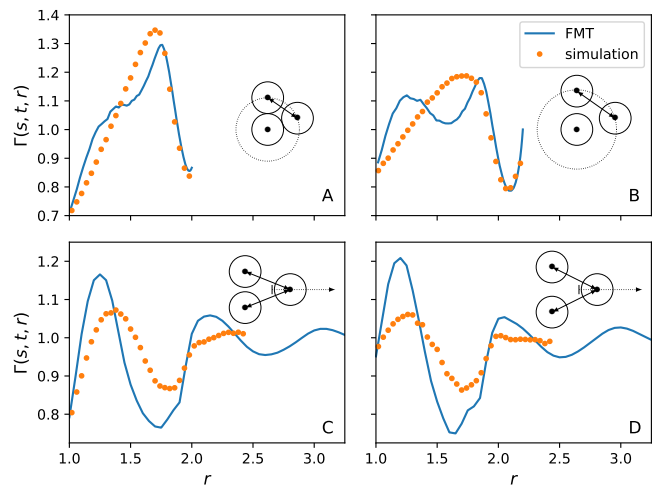

To investigate further this breakdown at high densities we compare in Fig.\ref{fig: triplet} the 
predictions of FMT with simulation data for a more varied selection of configurations at the 
even higher density, $\rho_b\!=\!0.8$. 
Panel A shows the variation for a rolling contact configuration. 
An unphysical `shoulder', already visible in panel C of Fig.\ref{fig: triplet 2}, becomes more 
pronounced, although one could argue that the overall description remains acceptable. 
This shoulder feature becomes more prominent when considering a 
rolling configuration with slightly more separation between the particles, shown in panel B of 
Fig.\ref{fig: triplet}. 
Despite showing reasonable behaviour at larger separations, for $r\!<\!2$ the description of the 
simulation data is rather poor. 
Panels C and D focus on stretched isoceles triangle configurations, which also serve to 
expose deficiencies of the FMT. 
While it is apparent that the general trends of the simulation data are roughly captured, 
the amplitude of oscillation is significantly overestimated. 
It would appear that the FMT performs best for rolling contact situations but leaves much 
to be desired at intermediate particle separations, at least for densities $\rho_b>0.7$.
It seems to us that the overall level of agreement of the FMT predictions with the 
Monte-Carlo data is on a similar level to that of earlier theories of the triplet correlation, 
such as those of Haymet {\it et al.} \cite{haymet1981} and Barrat {\it et al.} 
\cite{barrat1987}. For some examples of this we refer the reader to Fig.9 of the paper by 
Bilstein and Kahl \cite{kahl1994}. 

\section{Perturbation theory}\label{theorybh}

The hard-sphere model is not sufficient to capture all of the phenomena exhibited by real fluids. 
An improved description can be achieved if we supplement the hard-sphere repulsion with 
an attractive component to the interaction potential, 
$u=u_{\rm hs} + u_{\rm att}$.
If the attraction if sufficiently weak and long ranged, then the following first-order perturbation 
theory provides a good approximation to the Helmholtz free energy 
\begin{align}\label{inhom_bh}
F_{\rm BH}[\,\rho\,] &= F_{\rm hs}[\,\rho\,] 
\\
&\hspace*{-1.2cm}+ 
\frac{1}{2}\!\int\! d\rv_1\! \int\! d\rv_2\, 
\rho(\rv_1)\rho(\rv_2)u^{\rm att}(r_{12})
\big( 1 + h_{\rm hs}(\rv_1,\rv_2;[\,\rho\,]) \big),
\notag
\end{align}
where the first term is the Helmholtz free energy of hard-spheres, including the ideal gas contribution. 
The density enters equation \eqref{inhom_bh} both explicitly, via the quadratic product 
in the integrand, and implicitly, via the functional dependence of the hard-sphere Helmholtz 
free energy and total correlation function. 
In bulk, equation \eqref{inhom_bh} reduces to the well-known first-order perturbation 
theory of Barker and Henderson \cite{bh_original1,bh_original2,bh_review}. 
For this reason the approximation \eqref{inhom_bh} has been called the Barker-Henderson (BH) functional. 

In Ref.\cite{tschopp1} we investigated the density obtained from numerical minimization 
of the BH grand potential (using equations \eqref{grand}, \eqref{EQomegaMinimial} and \eqref{inhom_bh}) and found 
excellent agreement with simulation data for several inhomogeneous situations. 
Our findings suggest that the BH functional provides a quantitatively accurate description of inhomogeneous 
fluids with hard-core repulsion and weak attraction. 
Despite these promising results, widespread application of the BH functional, as implemented in 
\cite{tschopp1}, is likely to be hindered by the numerical effort required to minimize the 
grand potential. 
The strategy adopted, which we will henceforth refer to as the BH-PY approach, was to use FMT to approximate 
the first (reference) term in \eqref{inhom_bh} and to obtain $h_{\rm hs}$ by iteratively solving the 
inhomogeneous PY theory (equations \eqref{oz} and \eqref{py}).
Clearly, using the numerical solution of an inhomogeneous integral equation theory as part of a 
self-consistent minimization 
scheme is computationally expensive, particularly when larger systems are required (e.g.~studies of the liquid-vapour 
interface). 
However, we are now in a position to improve this situation by incorporating 
the FMT total correlation function, rather than that from the PY approximation, into \eqref{inhom_bh}. 
This BH-FMT approach yields a huge reduction in computation time and thus opens the door to applications which 
would be practically impossible using BH-PY.

We now briefly summarize the steps required to minimize 
the BH-FMT grand potential. Although this proceeds in much the same way as discussed in \cite{tschopp1}, there is a subtle 
but very important difference to be observed when evaluating the derivative contribution to the one-body direct 
correlation function. 
To limit the length of the presentation we restrict attention to the case of planar geometry, although similar calculations 
in spherical geometry would pose no greater difficulty. 
To solve the Euler-Lagrange equation \eqref{euler}, we require the one-body direct correlation function \eqref{1derivative}.
Using \eqref{inhom_bh} to evaluate the derivative yields 
\begin{align}\label{listoffour}
c^{(1)} = c_{\rm hs}^{(1)} + c_{\rm smf}^{(1)} + 
c_{\rm corr}^{(1)} + c_{\rm der}^{(1)},
\end{align}
where $c^{(1)}_{\rm hs}$ is given by \eqref{c_onebody}. 
The remaining terms involve integrals over the attractive interaction:
\begin{align}
c_{\rm smf}^{(1)}(\rv_1) &=-\!\int\!\! d\rv_2\, \rho(\rv_2)\beta u^{\rm att}(r_{12}),
\label{c_smf}
\\
c_{\rm corr}^{(1)}(\rv_1) &=-\!\int\!\! d\rv_2\, \rho(\rv_2)\beta u^{\rm att}(r_{12})h_{\rm hs}(\rv_1,\rv_2),
\label{c_corr}
\\
c_{\rm der}^{(1)}(\rv_1)&=-\!\int\!\! d\rv_2\! \int\!\! d\rv_3 
\frac{\rho(\rv_2)\rho(\rv_3)\beta u^{\rm att}(r_{23})}{2 }\,
\label{c_deriv}
\frac{\delta h_{\rm hs}(\rv_2,\rv_3)}{\delta\rho(\rv_1)}.  
\end{align}
The first of these is easy to calculate. 
The second contribution can be evaluated by using our analytic formulae for $\bar{c}_{\rm hs}$ as 
input to the OZ equation \eqref{oz_planar} and then transforming the resulting $\bar{h}_{\rm hs}$ back to real-space.

The functional derivative in \eqref{c_deriv} 
can be reexpressed in terms of a derivative with respect to the one-dimensional density profile
\begin{align}
\frac{\delta h_{\rm hs}(\rv_1,\rv_2)}{\delta\rho(\rv)}
=\frac{1}{A}\frac{\delta h_{\rm hs}(z_1,z_2,r)}{\delta\rho(z)},
\end{align}
where $A$ is an (arbitrary) area perpendicular to the $z$-axis (eliminated 
when performing the integrals in \eqref{c_deriv}). 
Using finite differences the functional derivative becomes 
\begin{align}\label{planar_rewrite}
\frac{\delta h_{\rm hs}(\rv_1,\rv_2)}{\delta\rho(\rv)}
=
\lim_{\varepsilon\rightarrow 0}
\frac{
h^{\varepsilon z\!}_{\rm hs}(z_1,z_2,r)
- 
h_{\rm hs}(z_1,z_2,r)
}{A\varepsilon},
\end{align}
where $h^{\varepsilon z}_{\rm hs}$ is the total correlation function evaluated at  
the perturbed density $\rho^{}_z(z_3)=\rho(z_3) + \varepsilon\,\delta(z_3 - z)$. 
%
%
%
%
Substitution of the perturbed density into the transformed OZ equation \eqref{oz_planar} yields 
\begin{align}\label{oz_planar_epsilon}
&\!\!\!\overline{h}^{\, \varepsilon z\!}_{\rm hs}(z_1,z_2,k) = 
\int_{-\infty}^{\infty} \!\!\!dz_3 \, 
\overline{h}^{\, \varepsilon z\!}_{\rm hs}(z_1,z_3,k)
\rho(z_3)
\overline{c}^{\, (2),\varepsilon z\!}_{\rm hs}(z_3,z_2,k)
\notag\\
& +\;\overline{c}^{\, (2),\varepsilon z\!}_{\rm hs}(z_1,z_2,k)
\,+\, \varepsilon\, \overline{h}^{\, \varepsilon z\!}_{\rm hs}(z_1,z,k) \overline{c}^{\, (2),\varepsilon z\!}_{\rm hs}(z,z_2,k).
\end{align}
Before solving \eqref{oz_planar_epsilon} for $\overline{h}^{\, \varepsilon z\!}_{\rm hs}$ we
evaluate the perturbed function $\overline{c}^{\, (2),\varepsilon z\!}_{\rm hs}$ using our analytical results.  
%
Since the Hankel transformed two-body direct correlation function depends on the density only through the 
weighted densities in 
$\Phi^{''}_{\alpha \beta}$, the perturbed pair direct correlation function can be obtained simply by 
substituting perturbed weighted densities, $n^{\varepsilon z\!}_{\alpha}$, into $\Phi^{''}_{\alpha \beta}$.
In planar geometry the weighted densities can be expressed as
\begin{equation}
n_{\alpha}(z_1) = \int dz_2 \; \rho(z_2)\, \omega_{\alpha}(z_1-z_2), 
\end{equation}
where the one-dimensional weight functions are given by
\begin{align*}
\omega_{3}(z) &\!=\! \pi \left(R^2\!-\!z^2\right) \Theta\left(R\!-\!|z|\right), \hspace*{0.05cm}
\omega_{2}(z) \!=\! 2 \pi R \Theta\left(R-|z|\right), \\
\omega_{1}(z) &\!=\! \frac{\omega_{2}(z)}{4 \pi R}, \hspace*{2.5cm}
\omega_{0}(z) \!=\! \frac{\omega_{2}(z)}{4 \pi R^2}, \\
\omega_{\bold 2}(z) &\!=\! 2 \pi z \, \unit_{z} \Theta\left(R-|z|\right), \hspace*{0.7cm}
\omega_{\bold 1}(z) \!=\! \frac{\omega_{\bold 2}(z)}{4 \pi R}.
\end{align*}
The perturbed weighted densities are therefore given by
\begin{align}
n^{\varepsilon z\!}_{\alpha}(z_1) = 
n_{\alpha}(z_1) + \varepsilon \, \omega_{\alpha}(z_1-z).
\end{align}
In \cite{tschopp1} equation \eqref{oz_planar_epsilon} was solved 
using the PY closure \eqref{py} on the perturbed functions $h^{\varepsilon z}_{\rm hs}$ and $c^{\varepsilon z}_{\rm hs}$ 
and solving the resulting {\it nonlinear} integral equation. 
This is much more demanding than the FMT route proposed here.

\begin{figure}
\vspace*{-0.6cm}
\begin{minipage}[t]{1\linewidth}
\begin{tikzpicture}
\node[inner sep=0pt] (fig11) at (0,0)
    {\includegraphics[width=1\linewidth]{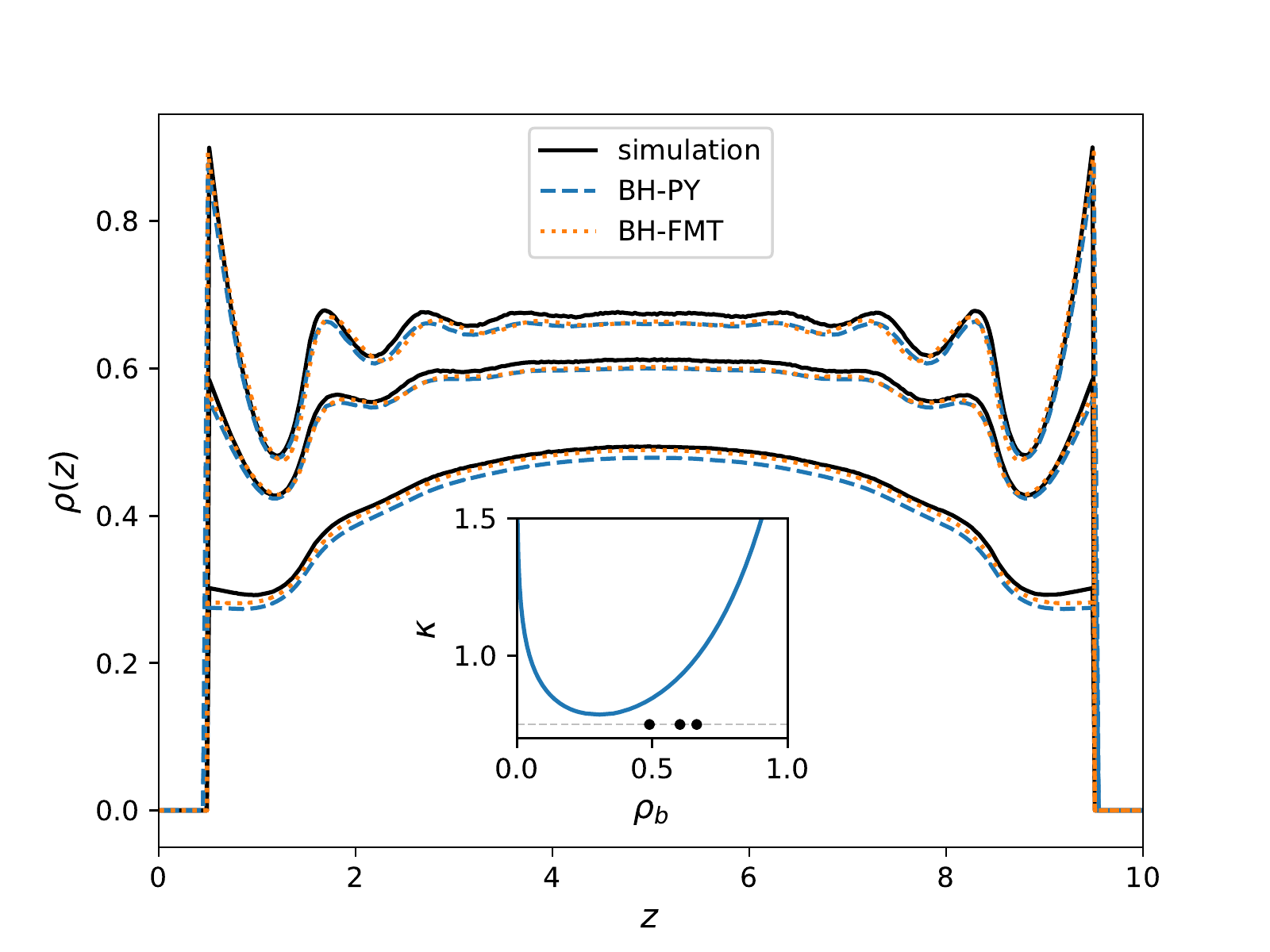}};
\draw[->, >=latex] (-0.5,-1.6925) -- (0.8,-1.6925);
\end{tikzpicture}
\end{minipage}
\caption{\textbf{Attractive hard-core Yukawa system under confinement.} 
Profiles for $\mu=-2.00$, $-1.50$ and $-1.00$, respectively, 
at parameter values $\kappa=0.75$ and $\alpha=1.8$. 
All profiles here are thus super-critical, recalling that $\kappa$ plays the role of an inverse temperature.
Simulation data (solid black lines) \cite{tschopp1}, 
BH-PY (dashed blue lines) and BH-FMT (dotted orange lines). The inset shows the phase diagram for $\alpha=1.8$, 
where we indicate the mid-point density of the calculated density profiles (black points) 
for fixed parameter $\kappa=0.75$ (dashed silver line).
}
\label{fig: slit}
\end{figure}

\section{Results for the hard-core Yukawa model}\label{resultsbh}

In Fig.\ref{fig: slit} we show density profiles for the attractive hard-core Yukawa (HCY) model 
confined between two planar walls separated by a distance of ten particle diameters. 
In addition to a hard-core repulsion the pair interaction potential of the HCY model has an 
attractive contribution given by
\begin{align}\label{hcy}
u_{\rm att}(r_{12}) = 
      -\kappa\,\frac{e^{-\alpha(r_{12}-1)}}{r_{12}} \hspace*{0.5cm} r_{12}\ge 1,
\end{align}
where $\kappa$ and $\alpha$ are positive constants.

Fig.\ref{fig: slit} shows profiles calculated using 
both the BH-PY and BH-FMT for three different chemical potentials. 
The theoretical predictions are compared with Monte-Carlo data taken from \cite{tschopp1}. 
The inset to Fig.\ref{fig: slit} shows the bulk binodal, where the points indicate the 
mid-point density of each of the considered profiles. 
For all three chemical potentials the BH-PY and BH-FMT are in very close agreement and both provide a good 
description of the simulation data.
We wish to emphasize that this close agreement between BH-FMT and BH-PY is a key result of the present 
work and central to the ultimate success of the method. 
Indeed, establishing this agreement provided much of the motivation for the present study of the two-body inhomogeneous 
correlation functions of hard-spheres. 
The fact that the accurate first-principles predictions of the BH-PY theory can be essentially 
reproduced, but with greatly reduced computational effort, by the BH-FMT is a significant step in 
turning the BH perturbation theory into a practically viable method for predicting the properties of 
realistic inhomogeneous fluids. 
While it remains to be seen whether the high level of agreement between BH-FMT and BH-PY remains 
in other situations, the data shown in Fig.\ref{fig: slit} seem to us to be very promising. 
To give the reader some feeling for the demands involved and the time saved by employing 
parallel computation - 
each of the BH-FMT profiles shown in Fig.\ref{fig: slit} required around four hours of computation 
time on a standard eight-core desktop machine (runtime effectively scales with the number of cores), 
whereas the corresponding BH-PY results each required several days to converge.

\section{Discussion}\label{discussion}

In this paper we have provided a detailed analysis of the inhomogeneous 
two-body correlation functions generated by FMT. 
Our formulae for the Hankel and Legendre transforms of the two-body 
direct correlation function enable rapid numerical evaluation of the real-space total correlation 
function and circumvent many of the usual numerical difficulties associated with iterative solution 
of inhomogeneous integral equation closures. 

Considering hard-spheres, our developments both facilitate the study of 
inhomogeneous microstructure and provide a fresh line of enquiry when analyzing the FMT. 
Past optimization strategies have focussed on thermodynamic (zero-body) and 
one-body quantities. 
It is our view that explicit consideration of the two-body correlation functions could lead to new 
insight into FMT, yielding
both quantitative criteria for the assessment of existing approximations as well as suggesting possible 
improvements. For example, it would be interesting to know the influence of either improved thermodynamics 
or tensorial weight functions on the predictions for the bulk triplet correlation function \cite{roth2010}.  
 
Using the FMT total correlation function as input to the BH perturbation theory \eqref{inhom_bh} 
yields a computationally viable approach for models of fluids with attractive interparticle 
interactions. 
The key advantage is that within FMT the inhomogeneous two-body correlation functions 
can be obtained using parallel computation. 
This is an essential feature if density functional theory beyond the one-body 
level is ever to become a practical tool for the investigation of relevant and interesting phenomena. 
The BH-FMT approach is undoubtedly much more efficient to implement than the BH-PY theory \cite{tschopp1} 
and does not appear to lead to any significant reduction in accuracy (see Fig.\ref{fig: slit}), even at high density. 
\\

At first sight, this conclusion might seem to be in contradiction to the triplet correlation data presented 
in Figs.\ref{fig: triplet 2} and \ref{fig: triplet}, where we find generally unsatisfactory performance of 
FMT at higher densities. 
We thus make the the following observations: 
(i) The integrand appearing in the BH free energy functional \eqref{inhom_bh} is weighted by the 
attractive part of the pair interaction potential, which lends particular importance 
to configurations around $r_{12}\!\approx\!1$. 
(ii) The integral in \eqref{inhom_bh} runs over both $\rv_1$ and $\rv_2$ and represents a complicated 
average over the inhomogeneous total correlation function.
These two features apparently lend the BH-FMT approach a certain robustness with respect to errors in the 
FMT hard-sphere two-body correlations and therefore open the door to many applications of the BH-FMT approach. 
However, only a more extensive investigation 
for different external fields and values of the model parameters will reveal under which conditions 
BH-PY and BH-FMT remain in such good agreement.

\appendix

\section{}\label{appendix_Phi''}

The nonzero derivatives of the free energy density, 
$\Phi^{'}_{\alpha}\!=\!\partial \Phi/\partial n_{\alpha}$, required for calculation of the FMT one-body 
direct correlation function are given by
\begin{align*}
\Phi^{'}_{0} &= -\ln(1-n_3) ,\;\;\; 
\Phi^{'}_{1} = \frac{n_2}{1-n_3} , \\
\Phi^{'}_{2} &= \frac{n_1}{1-n_3} + \frac{3 n_2^2- 3 \vec{n}_2 \cdot \vec{n}_2}{24 \pi (1-n_3)^2} , \\
\Phi^{'}_{3} &= \frac{n_0}{1-n_3} + \frac{n_1 n_2 - \vec{n}_1 \cdot \vec{n}_2}{(1-n_3)^2} + \frac{n_2^3 - 3 n_2 \vec{n}_2 \cdot \vec{n}_2}{12 \pi (1-n_3)^3} , \\
\Phi^{'}_{\bold 1} &= - \frac{\vec{n}_2}{1-n_3} ,\;\;\;\;\;
\Phi^{'}_{\bold 2} = - \frac{\vec{n}_1}{1-n_3} - \frac{n_2 \vec{n}_2}{4 \pi (1-n_3)^2} .
\end{align*}
The nonzero second derivatives, $\Phi^{''}_{\alpha \beta} \!=\! \partial^2 \Phi/\partial n_{\alpha}\partial n_{\beta}$, 
required to calculate 
the two-body direct correlation function are given by
\vspace*{-0.05cm}
\begin{align*}
\Phi^{''}_{0 3} &= \Phi^{''}_{3 0} = \frac{1}{1 - n_3},\;\;\;\;\;\;\;\;\;\;\;\;\; \Phi^{''}_{1 2} = \Phi^{''}_{2 1} = \frac{1}{1 - n_3} , \\
\Phi^{''}_{1 3} &= \Phi^{''}_{3 1} = \frac{n_2}{(1 - n_3)^2} ,\;\;\;\;\;\;\;\;\,\Phi^{''}_{2 2} = \frac{n_2}{4 \pi (1-n_3)^2} , \\
\Phi^{''}_{2 3} &= \Phi^{''}_{3 2} = \frac{n_1}{(1-n_3)^2} + \frac{n_2^2 - \vec{n}_2 \cdot \vec{n}_2}{4 \pi (1-n_3)^3}, \\
\Phi^{''}_{2 {\bold 2}} &= \Phi^{''}_{{\bold 2} 2} = -\frac{\vec{n}_2}{4 \pi (1-n_3)^2} , \\
\Phi^{''}_{3 3} &= \frac{n_0}{(1-n_3)^2} + \frac{2(n_1 n_2 - \vec{n}_1 \cdot \vec{n}_2)}{(1-n_3)^3} + \frac{n_2^3 - 3 n_2 \vec{n}_2 \cdot \vec{n}_2}{4 \pi (1-n_3)^4}, \\
\Phi^{''}_{3 {\bold 1}} &= \Phi^{''}_{{\bold 1} 3} = -\frac{\vec{n}_2}{(1-n_3)^2} , \\
\Phi^{''}_{3 {\bold 2}} &= \Phi^{''}_{{\bold 2} 3} = -\frac{\vec{n}_1}{(1-n_3)^2} - \frac{n_2 \vec{n}_2}{2 \pi (1-n_3)^3} , \\
\Phi^{''}_{{\bold 1} {\bold 2}} &= \Phi^{''}_{{\bold 2} {\bold 1}} = -\frac{1}{1-n_3} {\mathbb{1}} , 
\;\;\;\;\;\;\;\Phi^{''}_{{\bold 2} {\bold 2}} = -\frac{n_2}{4 \pi (1-n_3)^2} {\mathbb{1}} ,
\end{align*}
where ${\mathbb{1}}$ is the unit tensor.



\section{}\label{appendix_OZ}

We show how to obtain the Hankel transformed OZ equation for planar geometry 
\eqref{oz_planar} starting from the general expression \eqref{oz}. 
To clarify the treatment of `class 3' terms in the main text we find it 
convenient to break the calculation into two steps.

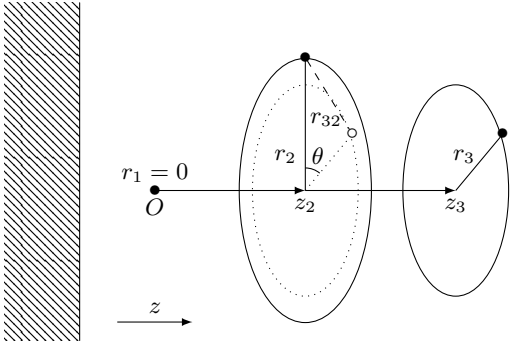
\begin{figure}
\begin{minipage}[t]{0.4\textwidth}
\hspace*{-0.5cm}
\begin{tikzpicture}
\coordinate (particle 1) at (0,0);
\coordinate (particle 2) at (2,1.77);
\coordinate (particle 3) at (4.625,0.75);
\coordinate (projected particle 3) at (2.625,0.75);
\coordinate (z2) at (2,0);
\coordinate (z3) at (4,0);
\coordinate (r2 label) at (2,0.45);
\coordinate (r3 label) at (4.375,0.45);
\coordinate (r32 label) at (1.95,0.95);
\coordinate (z axes label) at (0,-1.75);
\coordinate (z arrow left) at (-0.5,-1.75);
\coordinate (z arrow right) at (0.5,-1.75);
\coordinate (wall upper pt) at (-1,2.5);
\coordinate (wall lower pt) at (-1,-2);
\coordinate (wall size) at (-1,4.5);
\draw (z axes label) node[above] {$z$};
\draw (particle 1) node[below] {$O$};
\draw (z2) node[below] {$z_2$};
\draw (z3) node[below] {$z_3$};
\draw (particle 1) node[above] {$r_1=0$};
\draw (r2 label) node[left] {$r_2$};
\draw (r3 label) node[left] {$r_3$};
\draw (r32 label) node[right] {$r_{32}$};
\draw (particle 1) node {$\bullet$};
\draw (particle 2) node {$\bullet$};
\draw (particle 3) node {$\bullet$};
\draw[-, >=latex] (particle 2) -- (z2);
\draw[-, >=latex] (particle 3) -- (z3);
\draw[->, >=latex] (particle 1) -- (z2);
\draw[->, >=latex] (z2) -- (z3);
\draw (z2) ellipse (25pt and 50pt);
\draw (z3) ellipse (20pt and 40pt);
\draw[-, >=latex] (wall lower pt) -- (wall upper pt);
\fill[pattern=north west lines] (wall lower pt) rectangle ++(wall size);
\draw[->, >=latex] (z arrow left) -- (z arrow right);
\draw[dotted] (z2) ellipse (20pt and 40pt);
\draw[dotted, >=latex] (z2) -- (projected particle 3);
\draw[dashed, >=latex] (projected particle 3) -- (particle 2);
\pic [draw, -, angle radius=3mm, angle eccentricity=1.65, "$\theta$"] {angle = projected particle 3--z2--particle 2};
\draw (projected particle 3) node[white] {$\bullet$};
\draw (projected particle 3) node {$\circ$};
\end{tikzpicture}
\end{minipage}
\hspace*{2cm}
\caption{Sketch of the geometry used to rewrite the OZ equation in planar geometry. 
The point $O$ indicates the origin of the cylindrical coordinate system ($z_1=0, r_1=0$).}
\label{costheta_sketch}
\end{figure}

\textbf{Step 1:} 
When the density has a planar geometry we can express the OZ equation \eqref{oz} in the 
cylindrical coordinate system
\begin{align}\label{oz_cylindrical}
h(z_1, z_2, r_2) = c^{(2)}(z_1, z_2, r_2) + \int_{-\infty}^{\infty} dz_3 \int_{0}^{2 \pi}\!\! d\theta \\
\times \int_{0}^{\infty} dr_3 r_3\, h(z_1, z_3, r_3)\, \rho(z_3)\, c^{(2)}(z_3, z_2, r_{32}).
\notag
\end{align}
%
In Fig.\ref{costheta_sketch} we specify the geometry. 
If we choose $z_1$ as the axis of our cylindrical coordinates, then the $z$-projected separation between 
the points at $\rv_2$ and $\rv_3$ is given by $r_{32} = \sqrt{r_2^2 + r_3^2 - 2 r_2 r_3 \cos(\theta)}$.  
The Hankel back-transform of the pair direct correlation 
function \eqref{hankel_back} and the total correlation function can be expressed as
%
\begin{align}\label{back1}
c^{(2)}(z_1,z_2,r) 
&= 
\frac{1}{(2\pi)^2}\int \!d\vec{k}\, J_0(kr)\,\overline{c}^{\,(2)}(z_1,z_2,k), 
\\
h(z_1,z_2,r) 
&= 
\frac{1}{(2\pi)^2}\int \!d\vec{k}\, J_0(kr)\,\overline{h}(z_1,z_2,k), 
\label{back2}
\end{align}
where $d\vec{k}$ is an area element in the plane othogonal to the $z$-axis. 
Substitution of \eqref{back1} and \eqref{back2} into \eqref{oz_cylindrical} yields
\begin{align}\label{oz_intermediate}
&h(z_1, z_2, r_2) = 
\\
&c^{(2)}(z_1, z_2, r_2) + \frac{1}{(2 \pi)^4}\!\!\int_{-\infty}^{\infty} \!\!\!dz_3 \,\rho(z_3) \int_{0}^{\infty}\!\!\! 
dr_3 \,r_3  \int \!d\vec{k'}\!\! \int \!d\vec{k''} 
\notag\\
&\times \overline{h}(z_1, z_3, k')\, \overline{c}^{\,(2)}(z_3, z_2, k'') 
\,J_0(k'r_3) \!\!\int_{0}^{2 \pi} \!\!d\theta  J_0(k''r_{32}),
\notag
\end{align}
where we note that the separation $r_{32}$ is a function of $\theta$.
Graf's addition theorem for Bessel functions states that 
\begin{equation}
J_0(r_{23}) = \sum_{n=-\infty}^{\infty} J_n(r_2) J_n(r_3) e^{in\theta},
\end{equation}
which implies the useful result
\begin{equation}\label{grafintegral}
\int_{0}^{2 \pi} d\theta \; J_0(r_{23}) = 2 \pi J_0(r_2) J_0(r_3).
\end{equation}
Using \eqref{grafintegral} to perform the $\theta$-integral in \eqref{oz_intermediate} yields 
\begin{align}
h(z_1, z_2, r_2) =  c^{(2)}(z_1, z_2, r_2) + \int_{-\infty}^{\infty} \!\!dz_3\, \rho(z_3) 
\notag\\
\times \!\int d\vec{k'}\!\! \int \!d\vec{k''}\, \overline{h}(z_1, z_3, k')\, \overline{c}^{(2)}(z_3, z_2, k'')J_0(k''r_2) 
\notag\\
\times \frac{1}{(2 \pi)^3} \int_{0}^{\infty}\!\! dr_3\, r_3 J_0(k'r_3)  J_0(k''r_3).
\label{just_some_equation}
\end{align}
Bessel functions obey the orthogonality relation
\begin{align}
\label{orthogonality prop Hankel transf}
2 \pi \!\int_{0}^{\infty}\!\! dr_3 \; r_3 J_0(kr_3) J_0(k'r_3) = (2 \pi)^2 \delta(\vec{k}-\vec{k'}),
\end{align}
which could also be viewed as the Hankel transform of the zero-order Bessel function. 
Using this in \eqref{just_some_equation} yields
\begin{align}\label{step1final}
h(z_1, &z_2, r_2) = c^{(2)}(z_1, z_2, r_2) + \int_{-\infty}^{\infty} dz_3\, \rho(z_3) 
\\
&\times \frac{1}{(2 \pi)^2}\! \int d\vec{k'}\, \overline{h}(z_1, z_3, k')\, \overline{c}^{(2)}(z_3, z_2, k') J_0(k'r_2).
\notag
\end{align}

\textbf{Step 2:} 
Now that we have reexpressed the integration over the internal coordinate $\rv_3$ we will Hankel transform \eqref{step1final} 
with respect to the external coordinate $r_2$. 
Applying the operator $2\pi\int_0^{\infty} dr_2\,r_2 J_0(k r_2)$ to both sides of the equation yields
\begin{align}
&\overline{h}(z_1, z_2, k) = \overline{c}^{(2)}(z_1, z_2, k) + 
\frac{1}{2 \pi}\int_{-\infty}^{\infty} dz_3\, \rho(z_3) 
\\
&\times\!\!\! \int d\vec{k'}\, \overline{h}(z_1, z_3, k')\, \overline{c}^{(2)}(z_3, z_2, k') 
\!\!\!\int_0^{\infty}\!\!\! dr_2\,r_2 J_0(k r_2) J_0(k'r_2).
\notag
\end{align}
Using once more the orthogonality relation \eqref{orthogonality prop Hankel transf} then leads directly to the Hankel 
transformed OZ equation \eqref{oz_planar} in the main text.


\section{}\label{Appendix_OZspherical}

We show here the calculation analogous to that in the preceding Appendix, but now for spherical 
geometry. 
Starting from the general expression \eqref{oz} we obtain the Legendre transformed OZ equation \eqref{oz_spherical}
(closely following the presentation of Refs.\cite{attard1} and \cite{attard_book}).  
To clarify the treatment of `class 3' terms in the main text we break the calculation into two steps.


\textbf{Step 1:}
The OZ equation \eqref{oz} can be rewritten as
\begin{align}\label{oz_spherical2}
h(r_1, r_2, x_2) &= c^{(2)}(r_1, r_2, x_2) + \int_{0}^{\infty} \!dr_3 \; r_3^2\, \rho(r_3) \\
&\times \int_{0}^{2\pi} \!d\phi_3 \int_{-1}^{1} \!dx_3\, h(r_1, r_3, x_3)\, c^{(2)}(r_3, r_2, x_{32}),
\notag
\end{align}
where we have chosen the $z$-axis of the spherical coordinate system to coincide with the vector $\rv_1$, 
which implies $\theta_1\!=\!\phi_1\!=\!0$. Without loss of generality we can also orient the coordinates 
such that $\rv_2$ lies in the $xz$-plane, such that $\phi_2=0$. 
Using the back-transform \eqref{legendre_back} to represent both the pair direct and total correlation functions yields
%
\begin{align}\label{oz_intermediate_sph}
&h(r_1, r_2, x_2) = c^{(2)}(r_1, r_2, x_2) + \sum_{i,j=0}^{\infty} \int_{0}^{\infty} \!dr_3 \; r_3^2\, \rho(r_3)
\notag\\
&\times \int_{0}^{2\pi} \!d\phi_3 \int_{-1}^{1} \!dx_3\, \hat{h}(r_1, r_3, i)\, \hat{c}^{(2)}(r_3, r_2, j) P_i(x_3) P_j(x_{32}).
\end{align}
For our chosen orientation of coordinate system the addition theorem for spherical harmonics states that 
\begin{align}\label{addthmsphharmo}
P_j(x_{32}) &= P_j(x_3) P_j(x_2) \\
&\quad + 2 \sum_{m=1}^{j} \frac{(i-m)!}{(j+m)!} P_j^m(x_3) P_j^m(x_2) \cos(m \phi_3). \notag
\end{align}
Substitution of \eqref{addthmsphharmo} into \eqref{oz_intermediate_sph} and performing the integration over $\phi_3$ 
yields 
\begin{align}\label{just_some_equation_sph}
&h(r_1, r_2, x_2) = c^{(2)}(r_1, r_2, x_2) + \sum_{i,j=0}^{\infty} \int_{0}^{\infty} \!dr_3 \; r_3^2\, \rho(r_3)
\notag\\
&\times 2\pi \int_{-1}^{1} \!dx_3\, \hat{h}(r_1, r_3, i)\, \hat{c}^{(2)}(r_3, r_2, j) P_i(x_3) P_j(x_3) P_j(x_2).
\end{align}
Legendre polynomials obey the orthogonality relation
\begin{equation}
\label{orthogonality prop Legendre transf}
\int_{-1}^{\,1} \!dx\; P_i(x) P_j(x) = \frac{2}{2i+1} \delta_{ij},
\end{equation}
where $\delta_{ij}$ is the Kronecker delta.
Using this in \eqref{just_some_equation_sph} yields
\begin{align}\label{step1final_sph}
&h(r_1, r_2, x_2) = c^{(2)}(r_1, r_2, x_2) + \sum_{j=0}^{\infty} \int_{0}^{\infty} \!dr_3 \; r_3^2\, \rho(r_3)
\notag\\
&\times \frac{4\pi}{2j+1} \hat{h}(r_1, r_3, j)\, \hat{c}^{(2)}(r_3, r_2, j) P_j(x_2).
\end{align}

\textbf{Step 2:} 
Legendre transform \eqref{step1final_sph} 
with respect to the external coordinate $x_2$.
Applying the operator $\frac{2n+1}{2}\!\int_{-1}^{\,1} dx_2\,P_n(x_2)$ to both sides of the equation yields
\begin{align}
&\hspace*{-0.35cm}\hat{h}(r_1, r_2, n) = \hat{c}^{(2)}(r_1, r_2, n) + \sum_{j=0}^{\infty} \int_{0}^{\infty} \!dr_3 \; r_3^2\, \rho(r_3) \,2\pi
\\
&\times \frac{2n+1}{2j+1}\, \hat{h}(r_1, r_3, j)\, \hat{c}^{(2)}(r_3, r_2, j) \int_{-1}^{\,1} dx_2\,P_n(x_2) P_j(x_2).
\notag
\end{align}
Using once more the orthogonality relation \eqref{orthogonality prop Legendre transf} then leads directly to the Legendre 
transformed OZ equation \eqref{oz_spherical} in the main text.

\bibliographystyle{apsrev4-2} 
\bibliography{paper2}


\end{document}